\documentclass{article}

\usepackage{authblk}

\usepackage{amssymb,amsbsy,amsmath,amsfonts,amscd}
\usepackage{latexsym,euscript,exscale,epic,eepic,epsfig}
\usepackage{float}
\usepackage{calc}
\usepackage{bm}
\usepackage{graphicx}
\usepackage{enumerate}
\usepackage[bookmarks=false]{hyperref}
\usepackage{color}
\usepackage{pstricks,pst-node,pst-text,pst-3d,pst-plot}
\usepackage[subnum]{cases}
\usepackage[top=2.5cm,bottom=2.5cm,left=2.5cm,right=2.5cm]{geometry}

\def \tr{\mathop{\rm tr }\nolimits}

\DeclareMathOperator*{\argmin}{arg\,min}
\def\doubleunderline#1{\underline{\underline{#1}}}
\def \omb{\bar{\omega}}

\usepackage{ntheorem}
\newtheorem*{remark-non}{Remark}

\usepackage[utf8]{inputenc}
\usepackage{verbatim}
\usepackage{listings}
\usepackage{fancyvrb}
\usepackage{algorithm}
\usepackage{algpseudocode}
\usepackage{acronym}
\usepackage{caption}
\usepackage{booktabs}
\usepackage{mathtools}
\usepackage{subfig}
\usepackage{multirow}
\usepackage{enumitem}
\usepackage[round]{natbib}

\title{Phase-field study of crack nucleation and propagation in elastic - perfectly plastic bodies}

\author[1]{Stella Brach}
\author[2]{Erwan Tanné}
\author[3]{Blaise Bourdin}
\author[1]{Kaushik Bhattacharya}

\affil[1]{\small Division of Engineering and Applied Science, California Institute of Technology, Pasadena, CA 91125, USA}
\affil[2]{\small Department of Mathematics, University of British Columbia, Vancouver, BC V6T 1Z2 Canada}
\affil[3]{\small Department of Mathematics, Louisiana State University, Baton Rouge, LA 70803, USA}
\date{}
\begin{document}
\maketitle

\bibliographystyle{plainnat}
\pdfoutput=1

\begin{abstract}
{Crack initiation and propagation in elastic - perfectly plastic bodies is studied in a phase-field or variational gradient damage formulation.    A rate-independent formulation that naturally couples elasticity, perfect plasticity and fracture is presented, and used to study crack initiation in notched specimens and crack propagation using a surfing boundary condition.   Both plane strain and plane stress are addressed.  It is shown that in plane strain, a plastic zone blunts the notch or crack tip which in turn inhibits crack nucleation and propagation.  Sufficient load causes the crack to nucleate or unpin, but the crack does so with a finite jump.  Therefore the propagation is intermittent or jerky leaving behind a rough surface.  In plane stress, failure proceeds with an intense shear zone ahead of the notch or crack tip and the fracture process is not complete.
}
\end{abstract}

%%%%%%%%%%%%%%%%%%%%%%%%%%%%%%%%%%%%%%%%%%%%%%%%
%%%%%%%%%%%%%%%%%%%%%%%%%%%%%%%%%%%%%%%%%%%%%%%%
\section{Introduction}

The variational fracture field approach following~\citep{Bourdin-2000,Bourdin-2008} has emerged as a powerful tool for the study of fracture in brittle materials.  This approach is based on a regularization of the variational formulation of brittle fracture by~\citep{Francfort-1998} following~\citep{Ambrosio-1990}.  The crack set is approximated by a diffuse region described by a smooth fracture field variable, and an energy functional that approximates (in the sense of Gamma convergence) the fracture energy functional is minimized subject to boundary conditions.  This framework has now been widely used in various situations and also described as phase-field approach and the gradient damage approach (see for example,~\citep{Pham-2011,Klinsmann-2015,Pham-2017,Zhang-2017}).

Very few materials are brittle, and therefore the study of elastic - plastic fracture is a subject with a rich history; see~\citep{h_book_79} for a comprehensive review.  Rice and his collaborators~\citep{Rice-1968,rs_jmps_78}) studied the stress and strain fields in the vicinity of a stationary crack, and used the knowledge of these fields to understand the role of plasticity in enhancing fracture toughness.  In recent years, a damage model going back to~\citep{g_jmps_77} and~\citep{nt_jmps_87} has been widely used to study failure in elastic - plastic materials.  There is an understanding that cracks are blunted by the formation of a plastic zone around the crack tip, that high triaxiality ahead of the crack tip leads to voids and the crack propagates by the coalescence of voids (see~\citep{bl_aam_10} and the citations there).

Given its success in describing brittle fracture and given the importance of ductile failure, it is natural that the variational fracture field approach be extended to elastic - plastic materials, and this has been proposed by 
\citep{Alessi-2014, Alessi-2015, Ambati-2015, deBorst-1999, Miehe-2015, Miehe-2016, Nedjar-2001, Reusch-2003,Alessi-2014, Alessi-2015,Tanne-2017a},  who combined the~\citep{Bourdin-2008} functional with a scaled plastic dissipation functional.  They studied various aspects of the model including stability and homogeneous crack initiation in a one-dimensional medium as well as a uniaxial tensile specimen.~\citep{Miehe-2015, Miehe-2016} use a similar formulation though it is framed in the context of balance laws, and study selected two dimensional problems in infinitesimal and finite deformation.~\citep{Ambati-2015} use a different functional where the plastic strain modifies the elastic stiffness of the material.

This paper adopts the formulation of~\citep{Alessi-2014, Alessi-2015}, and implements it numerically to study crack nucleation and crack propagation in Mode I in both plane strain and plane stress.  
Crack nucleation is studied in a specimen with a notch subjected to increasing displacement boundary condition while crack propagation is studied in a rectangular domain subjected to a surfing boundary condition.  When the yield strength is high (compared to the nucleation stress of brittle fracture), plastic yielding is limited to a very small region around the crack tip, and the nucleation and propagation are broadly similar to those in brittle materials.  This is consistent with the idea of small scale yielding.  

The behavior changes for smaller yield strength.  In plane strain, a plastic zone forms at the notch (respectively crack) tip thereby  preventing nucleation at the notch (respectively pinning the crack).  Eventually for a high enough value of applied load, the crack does nucleate (respectively gets unpinned).  It does so by jumping through a finite distance both during nucleation and propagation.  It gets pinned again by the plastic zone and the cycle repeats.  In other words, the observed crack propagation in plane strain is always intermittent leaving behind a rough fracture surface.  

The observation of jerky motion appears to be a departure from the arguments of~\citep{rs_jmps_78,h_book_79} and others.  They compare a pinned crack (in an infinite medium loaded at infinity) and one that is steadily propagating, and show that the former has a smaller incremental energy release rate than that of the latter.  They argue that if the pinned crack were to start propagating continuously with time, the energy release rate would increase monotonically thereby ensuring the incremental stability of this solution.  The results here show that there is another solution to the problem -- one where the crack jumps.  The setting of rate independent, perfect plasticity means that one can have multiple solutions.  Further incremental stability of a solution involving crack propagating continuously with time does not imply stability against a solution involving crack jumps.

The source of the intermittent crack growth is the fact that a region of intense hydrostatic (tensile) stress appears ahead of the pinned crack tip.  This would suggest that a daughter crack would nucleate ahead of the crack tip and then propagate backwards to join the main crack.  This happens instantaneously in the current rate-independent, perfectly plastic model and manifests itself as a crack jump.   This understanding is consistent with the recent work cited in~\citep{bl_aam_10} (see also,~\citep{Beremin-1981a, Beremin-1981b, Benzerga-2016, nt_jmps_87, petal_am_16}) who argue that voids would nucleate ahead of the crack tip and eventually coalesce.  Further, the experimental observations of~\citep{kg_jmps_89} clearly shows the appearance of a daughter crack ahead of the main crack followed by coalescence albeit in a dynamical setting.  Finally, the jerky motion results in a rough or dimpled crack surface and this is well established in fractography.

The situation in plane stress is different.  A plastic zone forms at the crack tip and gradually extend ahead of it.  Failure occurs by the formation of a region of intense shear due to the high deviatoric stress induced by the lack of confinement.  This shear is accompanied by damage, but the fracture process is not complete.  Note that the assumption of plane stress requires that the material be thin compared even to the process zone.  Therefore, it is natural to wonder whether one encounters such situation in reality.

While the numerical implementation and the mechanisms they reveal are inherently interesting, a long term motivation to use both the insights from these mechanisms and numerical method to study crack propagation in heterogeneous elastic plastic materials.  This is addressed in forthcoming work.

The paper is organized as follows.  The formulation and the numerical method are described in Section~\ref{sec:approach}, and a few useful results are recalled in Section~\ref{sec:prelim}.  Crack nucleation in a notched specimen is studied in Section~\ref{sec:nucleation} while crack propagation under surfing boundary conditions is described in Section~\ref{sec:propagation}.  

\textit{Notation}. Underlined and double-underlined characters respectively denote vectors and second-order tensors. Blackboard letters indicate fourth-order tensors; symbol $:$ denotes the double-dot product operator. A superposed dot denotes the time derivative.

%%%%%%%%%%%%%%%%%%%%%%%%%%%%%%%%%%%%%%%%%%%%
%%%%%%%%%%%%%%%%%%%%%%%%%%%%%%%%%%%%%%%%%%%%
\section{A variational phase-field approach}\label{sec:approach}
Consider a body occupying a domain $\Omega\subset\mathbb{R}^2$ in its reference configuration.  Let $\underline{u}$ denote the displacement field, $\doubleunderline{\epsilon}=(\nabla\underline{u}+\nabla\underline{u}^t)/2$ the strain, $\doubleunderline{\sigma}$ the stress and $\doubleunderline{\epsilon}_\text{p}$  the plastic strain.   

Assuming elastic/perfectly-plastic behavior, the stress is confined to the set of admissible stresses $\mathcal{F}=\{\doubleunderline{\sigma}\,\, | \,\, f(\doubleunderline{\sigma})\leq 0\}$ related to the yield function $f$.  By the normality rule, the admissible rate of change of plastic strain is $\mathcal{G}=\{\dot{\doubleunderline{\epsilon}}_\text{p}\,\, | \,\, \dot{\doubleunderline{\epsilon}}_\text{p}=\lambda\, \partial f/\partial\doubleunderline{\sigma}\,\,\text{with}\,\,\lambda\geq0\}$, and the plastic dissipation is 
\begin{equation}
\label{eqn:pi}
\pi\big(\dot{\doubleunderline{\epsilon}}_\text{p}\big)=\sup_{\doubleunderline{\sigma}\in \mathcal{F}}\big(\doubleunderline{\sigma}:\dot{\doubleunderline{\epsilon}}_\text{p}\big).
\end{equation}
Clearly $\pi$ is homogeneous of degree one corresponding to perfect plasticity.  In this work,  the material is assumed to obey to a von Mises strength criterion with yield function $f\big(\doubleunderline{\sigma}\big)$ and plastic dissipation $\pi\big(\doubleunderline{\dot{\epsilon}}_\text{p}\big)$ 
\begin{equation}
\label{eqn:vM}
f\big(\doubleunderline{\sigma}\big)= \sigma_\text{eq}-\sigma_0\,, \qquad
\pi\big(\doubleunderline{\dot{\epsilon}}_\text{p}\big)=
\left\{
\begin{aligned}
&\sigma_0 \dot{\epsilon}^\text{eq}_\text{p}  \qquad \text{if} \quad \tr \doubleunderline{\dot{\epsilon}}_\text{p}=0\\
&+\infty \hspace{2em} \text{otherwise} 
\end{aligned}
\right.
\end{equation}
with $\sigma_\text{eq}=\sqrt{(3/2)\doubleunderline{\sigma}_\text{d}:\doubleunderline{\sigma}_\text{d}}$ and $\dot{\epsilon}^\text{eq}_\text{p}=\sqrt{(2/3)\doubleunderline{\dot{\epsilon}}_\text{p}^\text{d}:\doubleunderline{\dot{\epsilon}}_\text{p}^\text{d}}$, where 
$\doubleunderline{\sigma}_\text{d}=\doubleunderline{\sigma}-(1/3)(\tr\doubleunderline{\sigma})\doubleunderline{1}$
and $\doubleunderline{\dot{\epsilon}}_\text{p}^\text{d}=\doubleunderline{\dot{\epsilon}}_\text{p}-(1/3)(\tr\doubleunderline{\dot{\epsilon}}_\text{p})\doubleunderline{1}$ are the deviatoric stress and strain-rate tensors, respectively, and $\sigma_0$ is the von Mises strength.

Following~\citep{Bourdin-2000, Bourdin-2008}, fracture is described by introducing a scalar regularized phase field $\alpha: \Omega \to [0,1]$ with $\alpha=0$ corresponding to the intact material and $\alpha=1$ to complete fracture.  This fracture field varies at a length-scale $\ell > 0$.

The fracture of elastic - perfectly plastic bodies subject to some boundary condition is given by the solution of a minimization problem~\citep{Alessi-2015}: 
\begin{equation}
\label{eqn:min_reg_elpl}
\left(\underline{u}^\ast,\doubleunderline{\epsilon}_\text{p}^\ast,\alpha^\ast\right)_\ell\,\,=\,\,\argmin_{\substack{\underline{u}\in\mathcal{K}_\text{u},\,\,\, \dot{\doubleunderline{\epsilon}}_\text{p}\in\mathcal{G},\\  \dot{\alpha}\geq 0}} \,\widehat{\mathcal{E}}_\ell(\underline{u},\doubleunderline{\epsilon}_\text{p},\alpha)
\end{equation}
where the energy functional 
\begin{equation}
\label{eqn:func_ep}
\widehat{\mathcal{E}}_\ell \big(\underline{u},\doubleunderline{\epsilon}_\text{p},\alpha\big):=
\int_{\Omega}\frac{1}{2}\,\,(\doubleunderline{\epsilon}-\doubleunderline{\epsilon}_\text{p}):a(\alpha)\mathbb{C}:(\doubleunderline{\epsilon}-\doubleunderline{\epsilon}_\text{p})\,\,d\Omega+\int_\Omega \frac{G_\text{c}}{4c_w}\left[\frac{w(\alpha)}{\ell}+\ell |\nabla\alpha |^2 \right]\,d\Omega+
\int_{\Omega}\,b(\alpha)\int_{0}^{\overline{t}}\pi\big(\dot{\doubleunderline{\epsilon}}_\text{p}\big)d t\, d\Omega .
\end{equation}
Here, $\mathcal{K}_\text{u}$ is the set of kinematically admissible displacement fields subject to the applied boundary condition,  $\mathbb C$ is the elastic modulus (taken to be homogeneous and isotropic in what follows) and $G_c$ is the critical energy release rate (which will be referred to as ``toughness'' with a slight abuse of terminology).    The function $a(\alpha)=\eta_\text{e}+(1-\alpha)^2$ ensures that $\alpha=0$ corresponds to the intact material and $\alpha=1$ to complete fracture with $\eta_\text{e}$ being a small residual stiffness. The latter, introduced for the sake of numerical convenience, has been proven not to have any significant influence on the obtained results~\citep{Bourdin-2014}.  In this work, $w(\alpha)=\alpha$ and $c_w=2/3$. Finally,  $b(\alpha) = (1-\alpha)^2$  ensures that the yield strength is independent of the fracture.

In order to non-dimensionalize the equations, the fracture problem is recast in terms of the quantities
\begin{equation}
\label{eqn:adim}
\widetilde{\mathbb{C}}=\frac{\mathbb{C}}{E_0}\,,
\qquad
\widetilde{G}_\text{c}=\frac{G_\text{c}}{E_0 L_0}\,,
\qquad
\widetilde{\ell}=\frac{\ell}{L_0}\,,
\qquad
\widetilde{\delta}=\frac{\delta}{L_0}\,,
\qquad
\text{dim}\,\widetilde{\Omega}=\frac{\text{dim}\,{\Omega}}{L_0}
\end{equation}
where $L_0$ is a characteristic length of the domain $\Omega$, $E_0$, $\nu_0$ and $G_\text{c0}$ are the typical values of the elastic moduli and of the toughness parameter of an elastic material commonly used in engineering practice, and $\sigma_{\text{c}0}$ is the correspondingly-computed nucleation stress. For the sake of notation, the superposed tilde in Eq.\,\eqref{eqn:adim} is dropped in the following.

The minimization in Eq.\,\eqref{eqn:min_reg_elpl} is not trivial, since the regularized total energy $\widehat{\mathcal{E}}_\ell$ is non-convex. Nevertheless, noticing that $\widehat{\mathcal{E}}_\ell$ is separately convex in $\underline{u}$, $\doubleunderline{\epsilon}_\text{p}$ and $\alpha$,  the minimization problem is solved by an  iterating on three different sub-problems until convergence is reached, and according to the following two  procedures:
\begin{itemize}
\item \texttt{AUP}: for each time-step $t^i$, the functional $\widehat{\mathcal{E}}_\ell$ is minimized with respect to any kinematically-admissible displacement $\underline{u}^i$ and plastically-admissible strain $\doubleunderline{\epsilon}_\text{p}^i$ while holding the damage variable $\alpha^{i-1}$ fixed.  Then, fixing $\underline{u}^i$ and  $\doubleunderline{\epsilon}_\text{p}^i$, the energy is minimized with respect to $\alpha^i$, under the irreversibility condition $\alpha^i \geq \alpha^{i-1}$. 
\item \texttt{UPA}: for each time-step $t^i$ and for a given displacement field $\underline{u}^{i-1}$, the functional $\widehat{\mathcal{E}}_\ell$ is minimized with respect to the plastic strain $\doubleunderline{\epsilon}_\text{p}^i$ and the damage variable $\alpha^i$, respectively subjected to the normality rule and to the irreversibility condition. Then, given $\doubleunderline{\epsilon}_\text{p}^i$ and  $\alpha^i$, a further minimization is performed on the displacement field  $\underline{u}^i$ under the kinematic admissibility constraint.
\end{itemize}

The problems are implemented using linear finite elements. 
The constrained minimization with respect to the phase-field variable $\alpha$ is implemented using the variational inequality solvers provided by \texttt{PETSc}~\citep{Balay-1997,Balay-2013a, Balay-2013b}.
The minimization with respect to displacement field $\underline{u}$ is a linear problem, solved by using preconditioned conjugated gradients. The minimization with respect to the plastic strain  $\doubleunderline{\epsilon}_\text{p}$ is reformulated as a constrained optimization via \texttt{SNLP} solvers\footnote{Available at \url{http://abs-5.me.washington.edu/snlp/} and at \url{https://bitbucket.org/bourdin/ snlp}.}. All computations are performed by means of the open source code \texttt{mef90}\footnote{Available at \url{https://www.bitbucket.org/bourdin/mef90-sieve}.}.

%%%%%%%%%%%%%%%%%%%%%%%%%%%%%%%%%%%%%%%%%%%%%%%%%%%%%
%%%%%%%%%%%%%%%%%%%%%%%%%%%%%%%%%%%%%%%%%%%%%%%%%%%%%
\section{Preliminaries} \label{sec:prelim}
\begin{figure}
\centering
    {\includegraphics[width=0.8\textwidth]{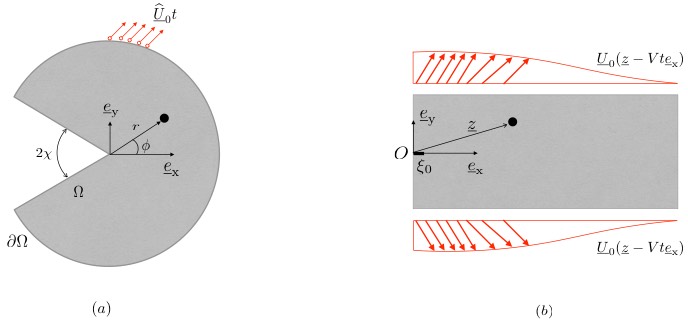}}
\caption{
{(a) Notch specimen undergoing a proportional loading used to study crack nucleation. (b)  Long specimen subject to surfing boundary conditions used to study crack propagation.}}
\label{fig:domain}
\end{figure}

The computational approach is used to study two problems shown in Figure~\ref{fig:domain}.  The first is a specimen with V-notch subjected to displacement boundary conditions to understand crack initiation and nucleation.  The second is a rectangular domain subjected to a surfing boundary condition to understand crack propagation.  In the figures in the sequel, the plastic process zone is shown in green and represents the region where the accumulated equivalent plastic strain 
$\epsilon_\text{p}^\text{eq}(\overline{t})=\int_0^{\overline{t}}\dot{\epsilon}_\text{p}^\text{eq}\,dt$ exceeds the value 0.1\%,  $\epsilon_\text{p}^\text{eq}(\overline{t})\geq 0.1\%$.  Similarly the presence of fracture process is shown by shading the region where the fracture variable exceeds the threshold 0.1\%,  $\alpha\geq 0.1\%$.   

The energy release rate is computed via the $J$-integral~\citep{Cherepanov-1967,Rice-1968} as
\begin{equation}
\label{eqn:J-integral}
J=\int_\mathcal{L} \left(\mathcal{W} n_{\mathcal{L}\text{x}} -\sigma_{ij}u_{i,\text{x}}n_{\mathcal{L}j}\right)\,ds 
\end{equation}
where $i,j\in\{x,y\}$, $\mathcal{W}$ is the elastic energy density, $\mathcal{L}$ is a contour surrounding the crack tip, and $\underline{n}_\mathcal{L}=n_{\mathcal{L}\text{x}}\underline{e}_\text{x}+n_{\mathcal{L}\text{y}}\underline{e}_\text{y}$ is the outward normal to $\mathcal{L}$.  The contour $\mathcal{L}$ is always chosen to be a outer boundary of the domain and away from any plastic region.

It is useful to recall two results from the linear elastic setting.  First, the toughness computed via the variational phase-field approach~\citep{Bourdin-2008} is increased by a factor that depends on the discretization $\delta$ so that the numerical critical energy release rate is
\begin{equation} \label{eq:gcnum}
G_\text{c}^\text{num}=G_\text{c} \left(1+ {\delta \over 4c_w\ell} \right).
\end{equation}
Second, crack nucleation depends on the regularization parameter $\ell$.  Specifically, in a uniaxial traction test, a crack  is nucleated when the local stress equals the critical threshold~\citep{Pham-2011}
\begin{equation}
\label{eqn:nucleation}
\sigma_{\text{c}}=\sqrt{\frac{3G_{\text{c}}E'}{8\ell}}\,,\qquad \text{with}
\qquad
E'=
\left\{
\begin{aligned}
&E  \hspace{4em}  \text{if plane stress,}\\
&\frac{E}{1-\nu^2} \hspace{1.9em} \text{if plane strain.} 
\end{aligned}
\right. 
\end{equation}
Indeed, this allows one to fit the regularization parameter to the known critical stress of the material.

In the simulations that follow, the yield strength is described relative to the nucleation stress using the {\it ductility ratio} 
\begin{equation} \label{eq:ry}
r_\text{y}={\sigma_\text{c} \over \sigma_0}.  
\end{equation}
As the yield strength $\sigma_0$ is reduced compared to the nucleation stress $\sigma_c$, the material can plastically deform before a crack is nucleated. As such, $r_\text{y}<1$ and $r_\text{y}>1$ respectively correspond to \textit{quasi-brittle} and \textit{ductile} fracture.

%%%%%%%%%%%%%%%%%%%%%%%%%%%%%%%%%%%%%%%%%%%%%%%%%%%%%
%%%%%%%%%%%%%%%%%%%%%%%%%%%%%%%%%%%%%%%%%%%%%%%%%%%%%
\section{Crack nucleation at stress singularities}\label{sec:nucleation}

This section studies the nucleation of cracks at a stress singularity following the approach of~\citep{Tanne-2017} who did so in the context of  elastic materials.

%%%%%%%%%%%%%%%%%%%%%%%%%%%%%%%%%%%%%%%%%%%%%%%%%%%%%
%%%%%%%%%%%%%%%%%%%%%%%%%%%%%%%%%%%%%%%%%%%%%%%%%%%%%
\subsection{Setting}
Consider a specimen occupying a domain $\Omega\subset\mathbb{R}^2$ consisting of a V-shaped notch, as shown in Figure~\ref{fig:domain}a, in its natural reference state. The notch angle is  $2 \omb $ with $\omb \in(0^+,\pi/2]$, such that the notch degenerates into a crack as $\omb \to0^+$ and opens up into a straight edge as $\omb \to\pi/2$.  The specimen is subjected to a proportional (time-scaling) displacement  $\underline{u}_0=\widehat{\underline{U}}_0 t$ along its exterior boundary $\partial\Omega$. By referring to a polar coordinate system $(r,\phi)$ emanating from the notch tip, the radial $\widehat{U}_{\text{r}}$ and tangential $\widehat{U}_{\phi}$ components of the displacement field $\widehat{\underline{U}}_0$ are taken to be
\begin{subequations}
\begin{align}
\widehat{U}_{\text{r}}&=
\left\{
\begin{aligned}
&\frac{r^\lambda}{E}\frac{F^{\prime\prime}(\phi)+(\lambda+1)(1-\nu\lambda)F(\phi)}{\lambda^2(\lambda+1)}  \hspace{17.6em}  \text{if plane stress}\\
&\frac{r^\lambda}{E}\frac{(1-\nu^2)F^{\prime\prime}(\phi)+(\lambda+1)[1-\nu\lambda-\nu^2(\lambda+1)]F(\phi)}{\lambda^2(\lambda+1)} \hspace{9em} \text{if plane strain} 
\end{aligned}
\right.
\\
\widehat{U}_{\phi}&=
\left\{
\begin{aligned}
&\frac{r^\lambda}{E}\frac{F^{\prime\prime\prime}(\phi)+[2(1+\nu)\lambda^2+(\lambda+1)(1-\nu\lambda)]F^\prime(\phi)}{\lambda^2(1-\lambda^2)}\hspace{10.7em}  \text{if plane stress}\\
&\frac{r^\lambda}{E}\frac{(1-\nu^2)F^{\prime\prime\prime}(\phi)+[2(1+\nu)\lambda^2+(\lambda+1)(1-\nu\lambda-\nu^2(\lambda+1))]F^{\prime}(\phi)}{\lambda^2(1-\lambda^2)}\qquad \text{if plane strain} 
\end{aligned}
\right.
\end{align}
\end{subequations}
where the prime denotes the derivatives with respect to $\phi$, the function 
\begin{equation}
F(\phi)=(2\pi)^{\lambda-1}\frac{\cos\left((1+\lambda)\phi\right)-g(\lambda,\omb )\cos\left((1-\lambda)\phi\right)}{1-g(\lambda,\omb )}\,,\quad
\text{with}
\quad
g(\lambda,\omb )=\frac{(1+\lambda)\sin\left((1+\lambda)(\pi-\omb )\right)}{(1-\lambda)\sin\left((1-\lambda)(\pi-\omb )\right)}
\end{equation}
and where $\lambda\in[0.5,1]$ is the exponent of the stress singularity at the crack tip computed by solving 
\begin{equation}
\label{eqn:lambda}
\sin\left(2\lambda(\pi-\omb ) \right)+\lambda\sin\left(2(\pi-\omb ) \right)=0.
\end{equation}
Note that this displacement field corresponds to the leading order term in the asymptotic field near notch subjected to a far-field mode-I conditions~\citep{Leguillon-1987}.  Thus, this boundary displacement promotes the nucleation of a mode-I crack at the notch tip.

Numerical simulations are performed using the  minimization procedure \texttt{AUP} in both plane stress and plane strain conditions.  The non-dimensional parameters are chosen as follows: the Young modulus $E=1.3$, Poisson ratio $\nu=0.3$, toughness $G_\text{c}=0.7$. The regularization length $\ell$ is chosen according to Eq.\,\eqref{eqn:nucleation} to make the nucleation stress  $\sigma_\text{c}=5$ and the mesh size $\delta=\ell/3$.  

\subsection{Crack initiation at a sharp notch}

\paragraph{Plane stress.}

\begin{figure}
\centering
    	{\includegraphics[width=1.0\textwidth]{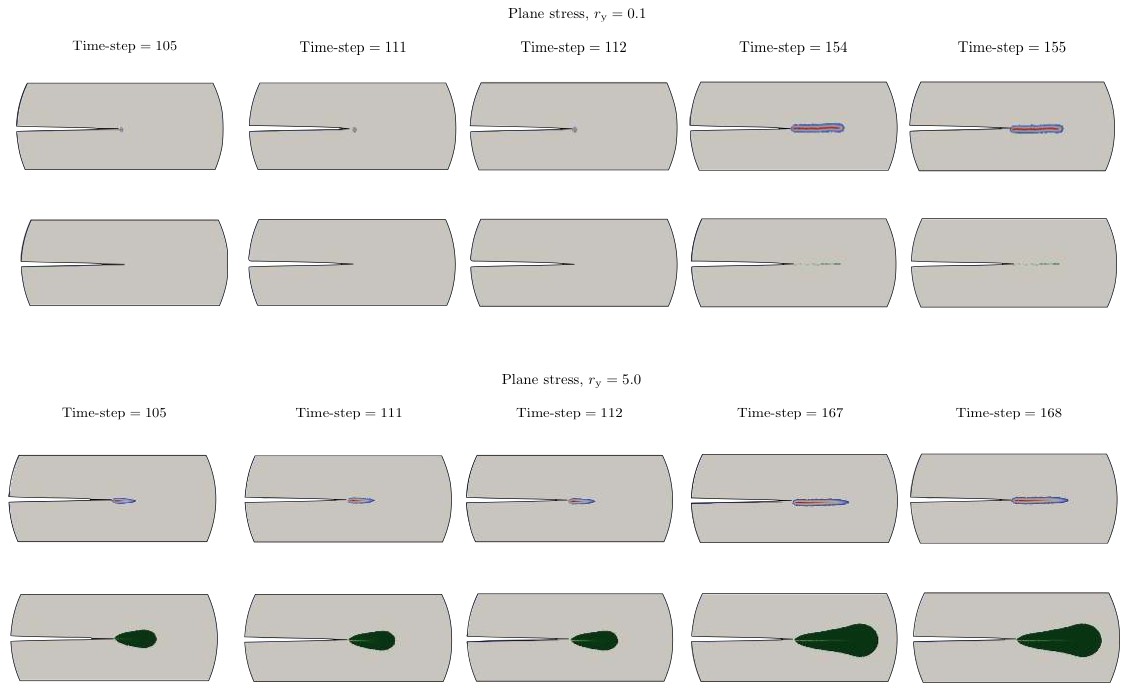}}
\caption{
{Crack initiation at a sharp notch in plane stress conditions. The fracture zone (top) and plastic process zone (bottom) for an elastic-plastic material  with $r_\text{y}=0.1$ and $r_\text{y}=5.0$.}}
\label{fig:Pac_nucl_plane_stress}
\end{figure}

\begin{figure}
\centering
   {\includegraphics[width=1.0\textwidth]{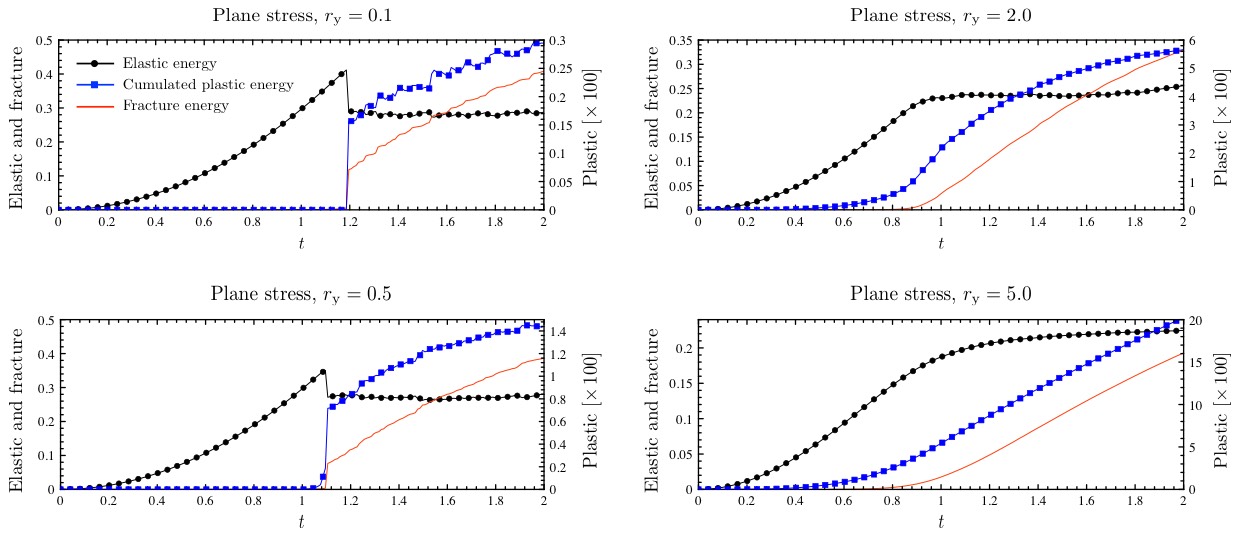}}
\caption{
{Crack initiation at a sharp notch in plane stress conditions. Various energies as a function of the time for an elastic-plastic material with $r_\text{y}=0.1$, $r_\text{y}=0.5$, $r_\text{y}=2.0$ and $r_\text{y}=5.0$.}}
\label{fig:Pac_en_plane_stress}
\end{figure}

\begin{figure}
\centering
   {\includegraphics[width=0.55\textwidth]{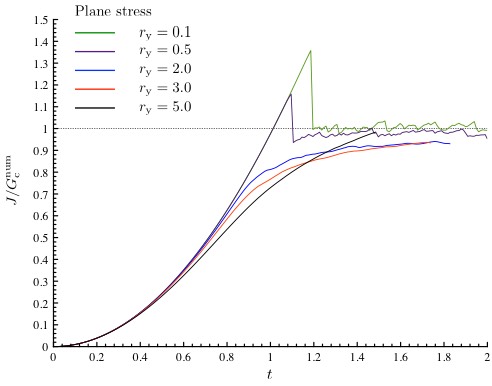}}
\caption{
{Crack initiation at a sharp notch in plane stress conditions. $J$-integral as a function of the time   for an elastic-plastic material  with different values of ductility ratio $r_\text{y}$.}}
\label{fig:Pac_J_plane_stress}
\end{figure}

\begin{figure}
\centering
   {\includegraphics[width=1.0\textwidth]{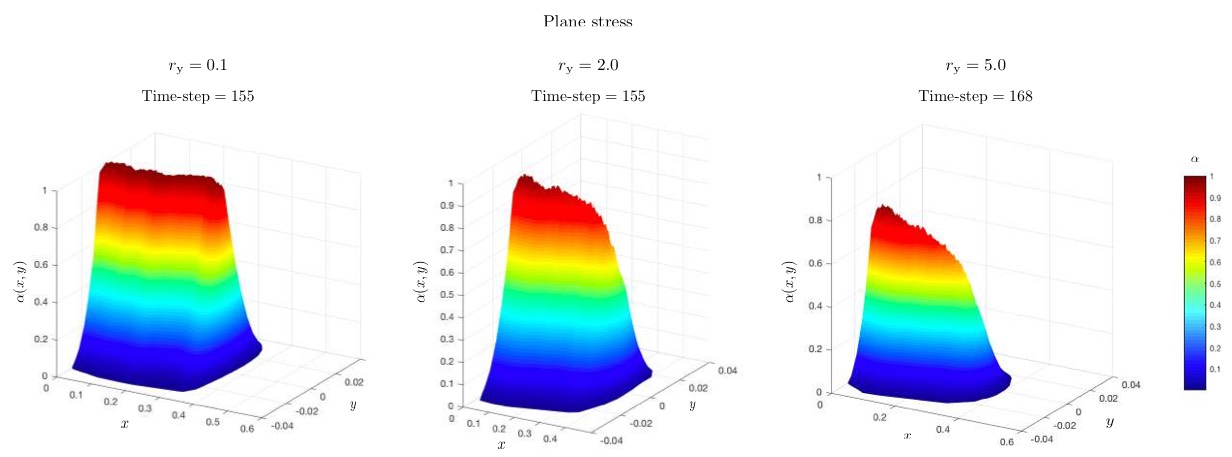}}
\caption{
{Crack initiation at a sharp notch in plane stress conditions. Fracture field computed for different values of the ductility ratio $r_\text{y}$.}}
\label{fig:Pac_J_plane_stress_damage}
\end{figure}

Figure~\ref{fig:Pac_nucl_plane_stress} shows the evolution of fracture and plastic fields for a sharp notch with $\omb =1^\circ$ in   {\it plane stress} conditions for various values of the ductility ratio $r_y$.  Figure~\ref{fig:Pac_en_plane_stress} shows the evolution of the various energies and Figure~\ref{fig:Pac_J_plane_stress} the energy release rate computed using the J-integral on the boundary.  Figure~\ref{fig:Pac_J_plane_stress_damage} shows the fracture field for various ductility ratio.

In the quasi-brittle situation $r_y = 0.1$ and $r_y=0.5$, the equivalent plastic strain $\epsilon_\text{p}^\text{eq}$ and fracture parameter $\alpha$ are zero outside a region of radius $O(\ell)$ at the notch tip at early times.  This is also reflected in the zero fracture energy and plastic dissipation, and the linear growth of the energy release rate.  At a critical time, a finite crack appears, accompanied by a sudden rise (respectively, reduction) of the fracture energy and plastic dissipation (respectively, elastic).  The crack then grows steadily, the plastic zone is confined to a small region around the crack tip and the energy release rate is close to the toughness $G_c$.  In other words, the behavior is very similar to the purely elastic or brittle situation~\citep{Tanne-2017}.

The behavior changes in the ductile situation $r_y = 2$ and $r_y=5$.  Plastic and fracture process zones appear soon after loading begins and grow steadily as the the load increases.   This is also evident in the smooth increase in fracture energy and plastic dissipation.  The energy release rate increases steadily and approaches the toughness, but does not reach that value.   In fact, the fracture variable does not reach the value 1 indicating that the fracture  process is not complete.  In other words, a band of intense plastic deformation forms and continues to grow along with progress towards but not complete fracture.  

There is also a curious cross-over in the energy release rate.  Note in Figure~\ref{fig:Pac_J_plane_stress} that at early stages, the energy release rate decreases with increasing ductility ratio $r_y$.  However, at some point, the energy release rate begins to reverse and eventually the energy release rate increases with increasing ductility ratio.  It is important to remember that the energy release rate computed 
using the J-integral on the boundary includes both the work of fracture and plastic dissipation.  Initially, the presence of plasticity impedes the fracture process and therefore the energy release rate is suppressed by ductility.  However, the plastic dissipation is higher in the more ductile materials, and this eventually dominates the energy release.

\begin{figure}
\centering
    {\includegraphics[width=1.0\textwidth]{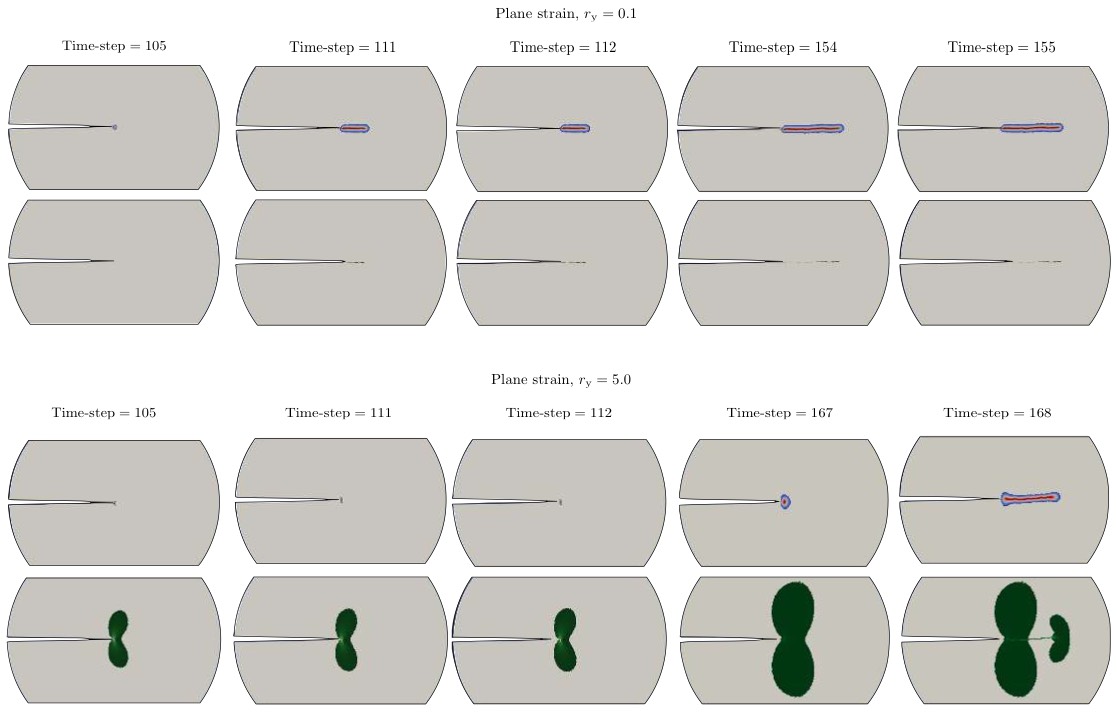}}
\caption{
{Crack initiation at a sharp notch in plane strain conditions. Plastic process zone and crack growth, for an elastic-plastic material  with  $r_\text{y}=0.1$ and $r_\text{y}=5$.}}
\label{fig:Pac_nucl_plane_strain}
\end{figure}

\begin{figure}[btp]
\centering
   {\includegraphics[width=1.0\textwidth]{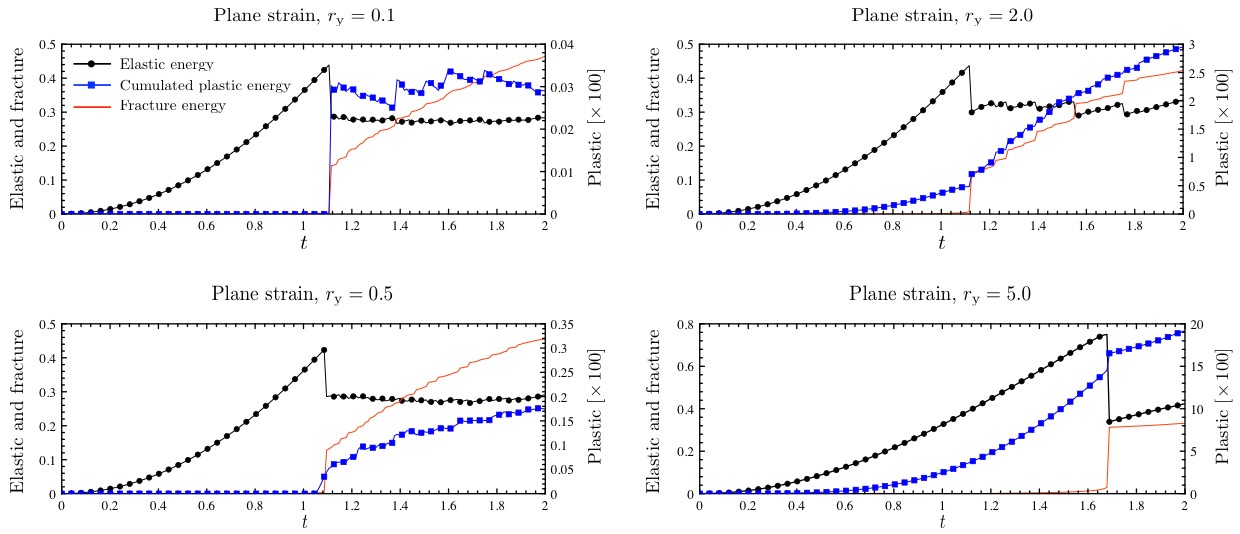}}
\caption{
{Crack initiation at a sharp notch in plane strain conditions. Various energies as a function of the time for an elastic-plastic material  with $r_\text{y}=0.1$, $r_\text{y}=0.5$, $r_\text{y}=2.0$ and $r_\text{y}=5.0$.}}
\label{fig:Pac_en_plane_strain}
\end{figure}

\begin{figure}[btp]
\centering
   {\includegraphics[width=0.55\textwidth]{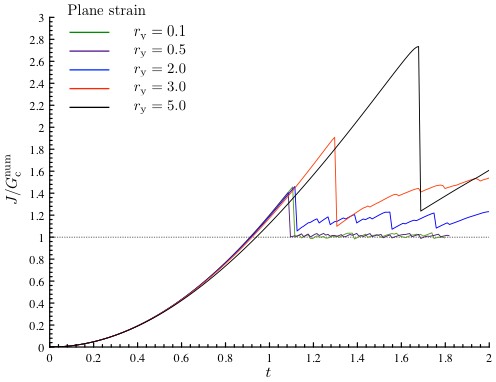}}
\caption{
{Crack initiation at a sharp notch in plane strain conditions.  $J$-integral as a function of the time   for an elastic-plastic material  with different values of $r_\text{y}$.}}
\label{fig:Pac_J_plane_strain}
\end{figure}

\begin{figure}
\centering
   {\includegraphics[width=1.0\textwidth]{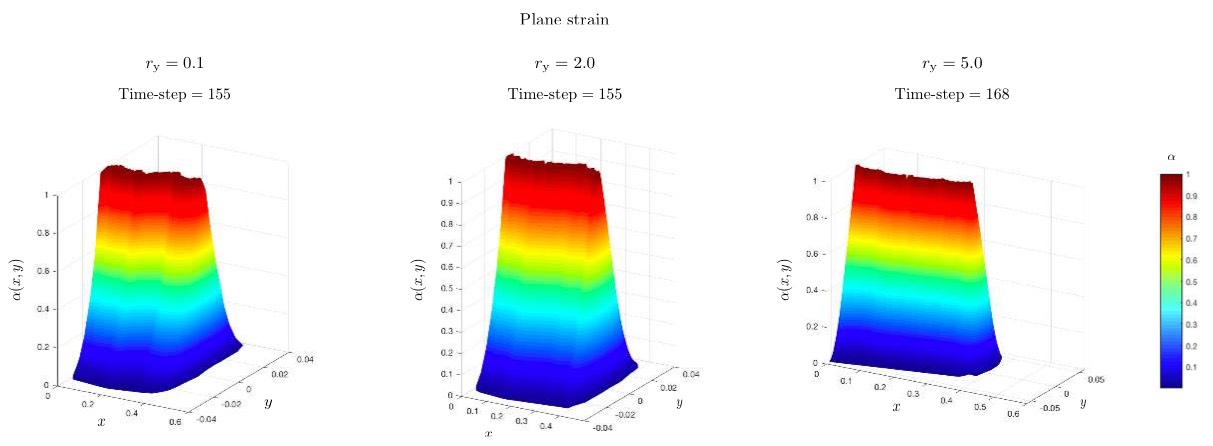}}
\caption{
{Crack initiation at a sharp notch in plane strain conditions. Fracture field computed for different values of the ductility ratio $r_\text{y}$.}}
\label{fig:Pac_J_plane_strain_damage}
\end{figure}

\paragraph{Plane strain.}

The same problem is studied in {\it plane strain} and the results are shown in Figures~\ref{fig:Pac_nucl_plane_strain},~\ref{fig:Pac_en_plane_strain},~\ref{fig:Pac_J_plane_strain} and~\ref{fig:Pac_J_plane_strain_damage}.  The behavior in the quasi-brittle situation is similar to that of plane stress and the purely elastic case.  However, the behavior in the ductile situation in plane strain is very different from that of plane stress.  In plane strain, a small region of non-zero fracture field variable and plastic deformation forms early in the loading.  The plastic zone continues to grow in two lobes either side of the mid-plane but the fracture field remains essentially fixed.   This continues till the energy release rate exceeds the toughness of the material as well as the critical value of the quasi-brittle case.  Then, at a critical load, the fracture field increases suddenly corresponding to the formation of a finite crack (note that the fracture field reaches the value one) and accompanied by a drop in the elastic strain energy.  Continued loading creates a plastic zone near the crack tip but keeps the crack pinned at the same location and the energy release rate again begins to grow.  This continues till the energy release rate reaches a different critical value when the crack suddenly jumps forward by a finite distance, and the elastic strain energy and energy release rate drop.  The cycle then repeats.  Importantly, note that the critical value of energy release rate required for crack initiation increases with ductility.

\paragraph{Summary.}
Crack initiation at a sharp notch in quasi-brittle materials is similar to that in purely elastic brittle materials.  However, the behavior changes for ductile materials.  Further, the behavior in ductile materials is quite different in plane stress and plane strain.  In the former, an intense plastic band accompanied by a fracture field region forms and grows smoothly ahead of the notch.   Further, the fracture process is not complete.  In contrast, in plane strain, the crack initiation and growth is impeded by plasticity and proceeds in an intermittent fashion.  Further, the fracture process proceeds to completion.  This difference reflects the difference in confinement.  In plane stress, the lack of confinement leads to high deviatoric stress ahead of the notch/crack.  This promotes plastic deformation which in turn promotes increase in the fracture field but not complete fracture.  In contrast, in plane strain, the confinement limits the deviatoric stress ahead of the notch/crack but promotes it away from it.  This results in a growth of the plastic zone that drains away energy from the fracture field.  Thus, the crack is pinned; however, once it breaks free it does not see any pinning and therefore jumps through a finite distance.

\newpage
\subsection{Crack nucleation at a V-notch}

\begin{figure}
\centering
	\subfloat[]
    {\includegraphics[width=1.0\textwidth]{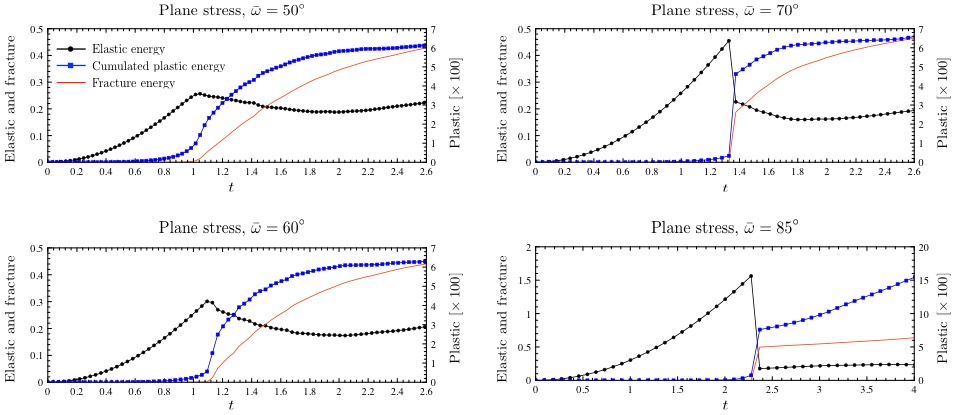}}\\
    \subfloat[]
    {\includegraphics[width=0.9\textwidth]{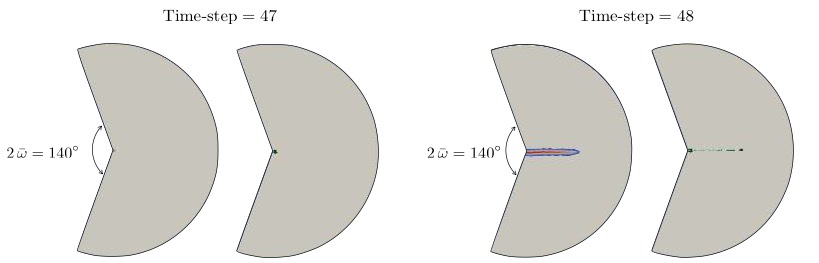}}\\
    \subfloat[]
    {\includegraphics[width=0.9\textwidth]{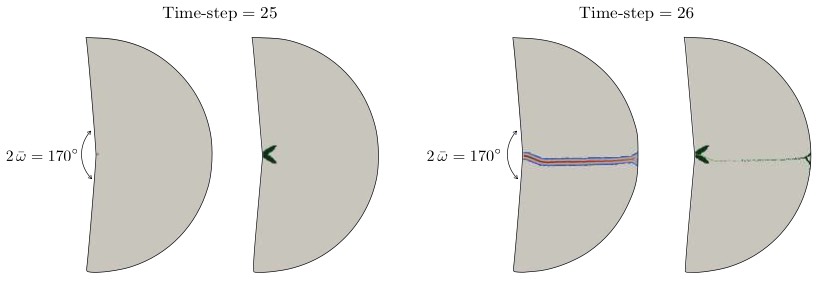}}
\caption{
{Crack nucleation at a V-notch in plane stress conditions for an elastic-plastic material with ductility ratio $r_y=2$. (a) Energies as a function of time $t$. (b) Crack nucleation and plastic process zone, for $\omb =70^\circ$. (b) Crack nucleation and plastic process zone for $\omb =85^\circ$.}}
\label{fig:Pac_angles_energies_plane_stress}
\end{figure}

\begin{figure}
\centering
   {\includegraphics[width=1.0\textwidth]{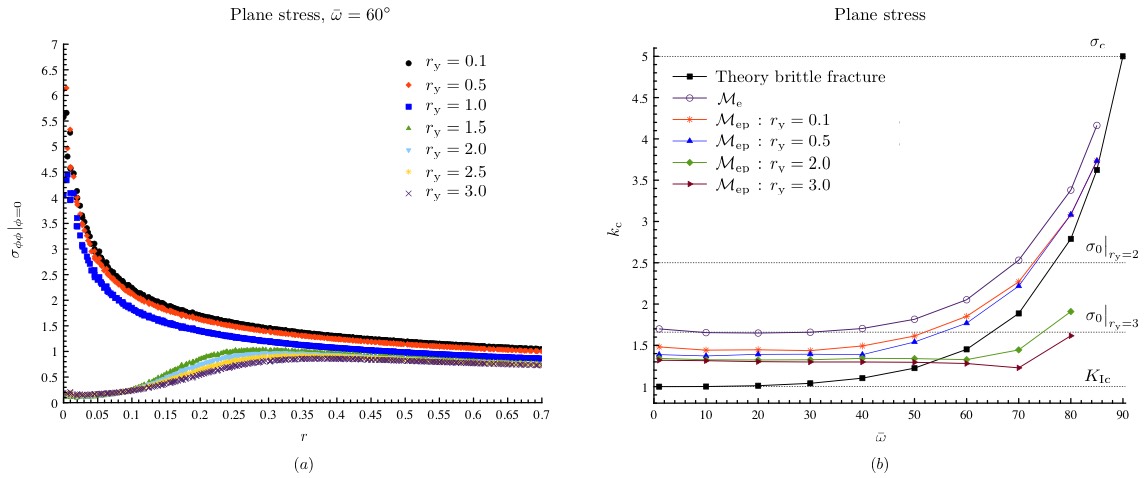}}
\caption{
{Crack nucleation at a V-notch in plane stress conditions.  (a) The azimuthal normal stress ahead of the crack tip for various ductility ratios.  (b) The critical value of the stress intensity as a function of notch opening angle for various ductility ratios.}}
\label{fig:plane_stress_notch_anal}
\end{figure}

A V-notch with a larger opening angle is considered next.  The quasi-brittle behavior in both plane stress and plane strain is similar to the elastic case studied by~\citep{Tanne-2017}, and not discussed any further.  In contrast, the ductile behavior is quite rich.

\paragraph{Plane stress.}

Figure~\ref{fig:Pac_angles_energies_plane_stress}(a) shows the evolution of the various energies for the case $\omb  = 50^\circ$, $\omb  = 60^\circ$, $\omb  = 70^\circ$  and $\omb  = 85^\circ$ in {\it plane stress}.  For small notch opening angles, the evolution is similar to the earlier case of a sharp notch -- i.e., an intense band of shear and fracture field appears ahead of the crack and grows smooth with increasing load.  However, the initial growth of the fracture zone becomes more and more sluggish with increasing opening angle.   This reflects that the stress concentration becomes weaker with increasing the notch angle.
Indeed, for $\omb  = 50^\circ$ and $\omb  = 60^\circ$, the fracture activity does not begin till a critical time.  It proceeds smoothly once it begins and is accompanied with a smooth drop of elastic energy.  The behavior  transitions to that of pinning followed by sudden growth for larger notch opening angle as shown in the case of $\omb =70^\circ$ in Figure~\ref{fig:Pac_angles_energies_plane_stress}(b).  In this case, the plastic and fracture zone is confined to the notch-tip till the load reaches a critical value when a finite fracture zone appears with a drop in elastic energy.  As in the previous cases, the fracture does not go to completion (the figure is not shown).  The fracture zone is pinned at further loading as a plastic zone appears and grows near the fracture tip.  The behavior transitions again for higher $\omb $. At $\omb =85^\circ$ when the notch opens to almost a straight edge, the fracture zone is no longer straight, but appears at a $45^\circ$ angle as shown in Figure~\ref{fig:Pac_angles_energies_plane_stress}(c).

%The reason behind this change of behavior with notch opening angle lies in the behavior of the stress field at the notch tip.  This is shown in Figure~\ref{fig:plane_stress_notch_anal}(a).  The ductility of the material regularizes the stress singularity at the notch tip.  Still, the stress is elevated for small notch angle.  However, around $\omb =50^\circ$, the stress drops at the notch tip and this delays the fracture activity. 

Crack nucleation at a notch was studied extensively by~\citep{Tanne-2017} in the brittle case.  They introduced a generalized stress-intensity factor:
\begin{equation}
\label{eqn:kfactor}
k=\left.\frac{\sigma_{\phi\phi}}{(2\pi r)^{\lambda-1}}\right |_{\phi=0}
\end{equation}
where $\lambda$ is the order of the singularity in the displacement field of an elastic field around a notch.  Since the hoop stress has a singularity of strength $\lambda -1$, this $k$ is well-defined independent of radius.  In particular, note that this $k$ equals the stress in the case of a straight edge and the stress-intensity factor in the case of a crack.~\citep{Tanne-2017} showed that a large amount of experimental and computational data in brittle fracture can be explained by postulating a crack nucleation criterion 
\begin{equation} \label{eq:k}
k=k_c = K_{Ic}^{2-2\lambda} \sigma_c^{2\lambda -1}
\end{equation}
where the critical value $k_c$ interpolates between $K_{Ic}$ and $\sigma_c$.  A similar criterion was also shown to hold at bimaterial interfaces~\citep{Hsueh-2018}.

Figure~\ref{fig:plane_stress_notch_anal}(b) studies the applicability of this criterion in the elastic plastic materials in plane stress.  This figure plots the critical value of $k$ computed as the value at the time-step just preceding the one where the fracture energy jumps or reaches the value $0.1$.  The behavior of the quasi-brittle case ($r_y < 1$) is similar to the elastic case and follows~(\ref{eq:k}).  However, the behavior is dramatically different for the ductile situations.  The reason for this can be found in Figure~\ref{fig:plane_stress_notch_anal}(a): the stress concentration at the notch tip disappears in the ductile case even for the sharpest notch.  Therefore, critical $k$ is limited by the value dictated by shear failure, and is relatively independent of the notch opening angle.

\begin{figure}
\centering
	\subfloat[]
    {\includegraphics[width=1.0\textwidth]{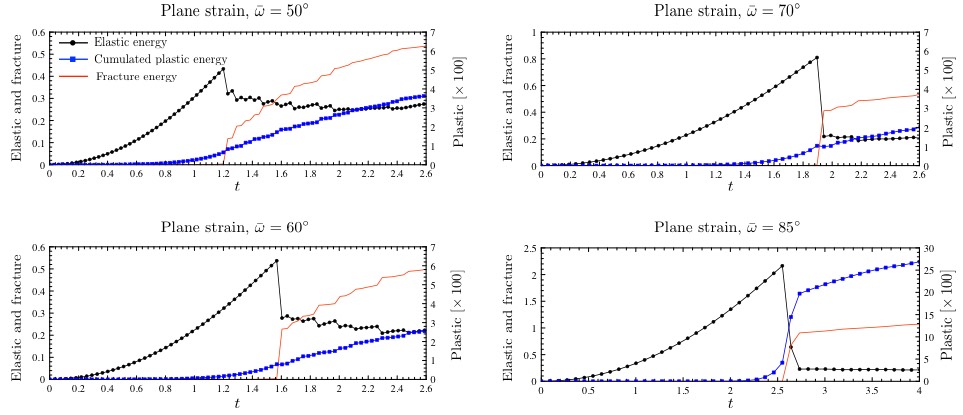}}\\
    \subfloat[]
    {\includegraphics[width=0.9\textwidth]{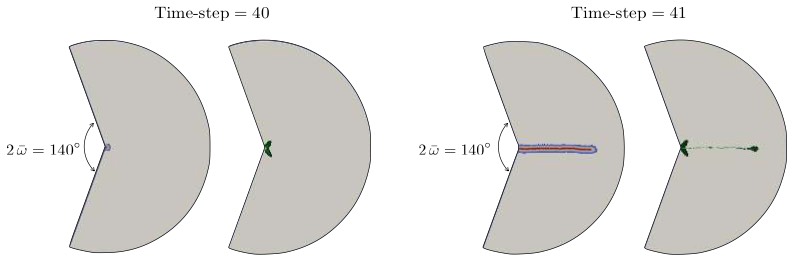}}\\
     \subfloat[]
     {\includegraphics[width=0.9\textwidth]{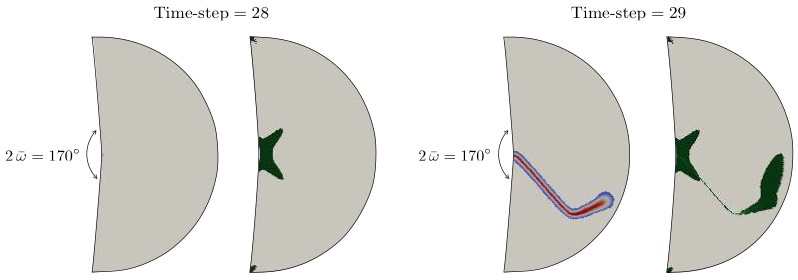}}\\
\caption{
{Crack nucleation at a V-notch in plane strain conditions for an elastic-plastic material with ductility ratio $r_y=2$. (a) Energies as a function of time $t$. (b) Crack nucleation and plastic process zone, for $\omb =70^\circ$. (b) Crack nucleation and plastic process zone for $\omb =85^\circ$.}}
\label{fig:Pac_angles_energies_plane_strain}
\end{figure}

\begin{figure}
\centering
	\subfloat[]
   {\includegraphics[width=1\textwidth]{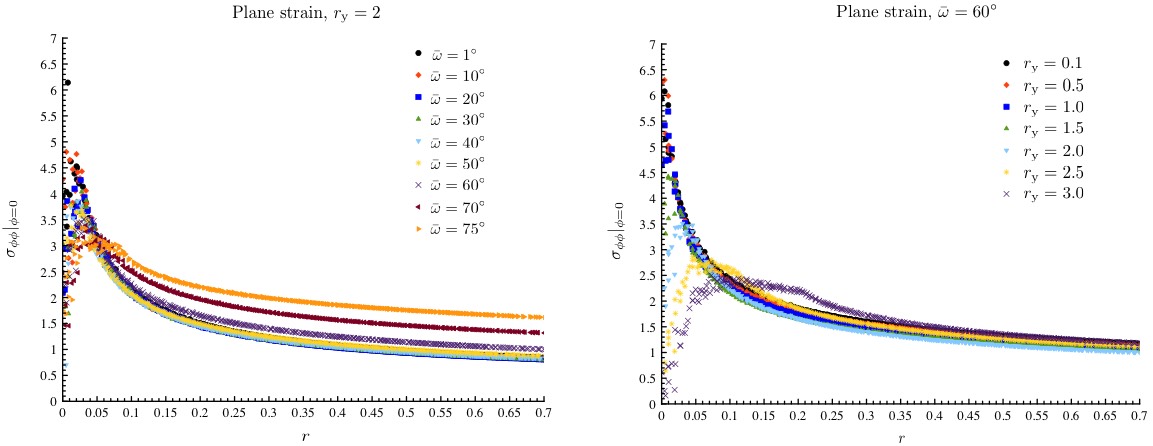}}\\
   \subfloat[]
   {\includegraphics[width=1\textwidth]{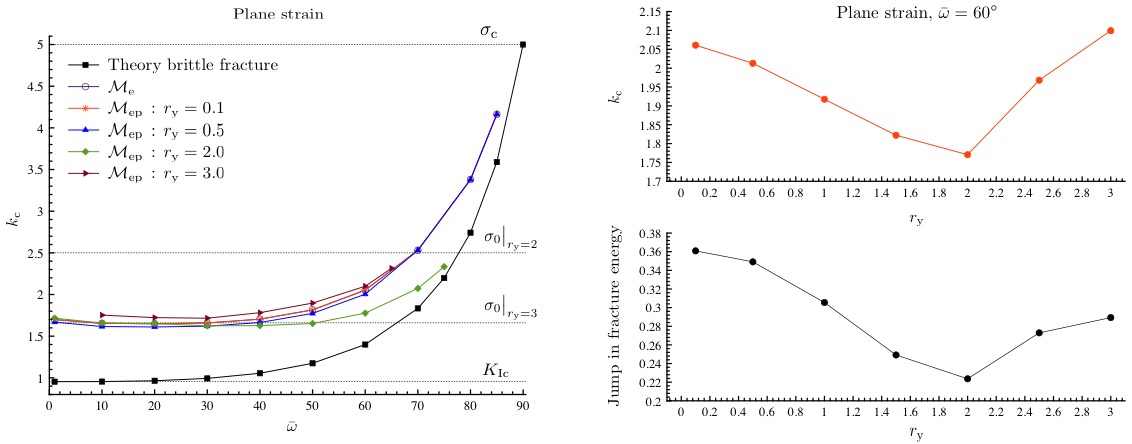}}
\caption{
{Crack nucleation at a V-notch in plane strain conditions. (a) The azimuthal normal stress ahead of the crack tip for various notch opening angle and ductility ratios.  (b) The critical value of the stress intensity as a function of notch opening angle for various ductility ratios.}}
\label{fig:plane_strain_notch_anal}
\end{figure}

\paragraph{Plane strain.}

The {\it plain strain} situation is shown in Figure~\ref{fig:Pac_angles_energies_plane_strain}(a).  As in the case of a sharp notch, there is a distinct nucleation event where a crack forms suddenly and with finite length.  Further, the fracture always proceeds to completion.  The plastic zone has the shape of two lobes just above and below the symmetry plane.  Finally, for high notch opening angles, as one approaches the straight edge, the crack forms at $45^\circ$ angle (Figure~\ref{fig:Pac_angles_energies_plane_strain}(c)).

The behavior of the stress near the notch tip is shown in Figure~\ref{fig:plane_strain_notch_anal}(a).  The ductility regularizes the stress field, but there is always a region of elevated stress in the vicinity of the notch tip.   This contributes to the nucleation of a finite (complete) crack in plane strain.  The critical value of the generalized stress intensity is shown in Figure~\ref{fig:plane_strain_notch_anal}(b).  The behavior of quasi-brittle cases are similar to those of the brittle case.  However, the behavior of the ductile cases are quite varied.  The critical value drops and becomes independent of $\omb $ for ductility ratio $r_y=2.0$, but begins to rise again with $\omb $ for higher values of ductility ratio.  This is highlighted for the case $\omb =60^\circ$ where  the behavior of $k_c$ vs. $r_y$ is non-monotone.    The stress distribution for this situation is on the right of Figure~\ref{fig:plane_strain_notch_anal}(a).  The plasticity regularizes the stress singularity, but a region of elevated stress remains and in fact spreads with increasing $r_y$.  This initially promotes crack nucleation till about $r_y=2.0$.  However, for higher ductility ratio the elevation in stress is minimal and this makes crack nucleation more difficult.  This manifests itself in the non-monotone behavior.

\paragraph{Summary.}
Crack nucleation at a V-notch in the quasi-brittle situation is very similar to that in purely elastic brittle materials.  However, the behavior of ductile materials is different and rich.  In plane stress, a region of intense plastic deformation accompanied by a fracture field forms and grows smoothly ahead of the notch for small notch angles.  For large notch angles, the growth of the fracture field is inhibited till a critical load, and then proceeds either rapidly or suddenly.  For very large notch angle, the plastic and fracture zone form at a $45^\circ$ angle.  In all these situations, the fracture never proceeds to completion.  In plane strain, a plastic zone forms near the notch tip that inhibits crack growth.  A finite crack nucleates suddenly and the fracture process is complete in this crack.  For notch opening angles, the critical generalized stress intensity when the crack forms behaves non-monotonically with ductility ratio due to the behavior of the stress field at the notch tip.

%%%%%%%%%%%%%%%%%%%%%%%%%%%%%%%%%%%%%%%%%%%%%%%%%%%%%
%%%%%%%%%%%%%%%%%%%%%%%%%%%%%%%%%%%%%%%%%%%%%%%%%%%%%
\section{Crack propagation}\label{sec:propagation}

This section studies the propagation of cracks subjected to surfing boundary conditions folllowing~\citep{Hossain-2014}.

%%%%%%%%%%%%%%%%%%%%%%%%%%%%%%%%%%%%%%%%%%%%%%%%%%%%%
%%%%%%%%%%%%%%%%%%%%%%%%%%%%%%%%%%%%%%%%%%%%%%%%%%%%%
\subsection{Setting}

Consider the rectangular domain $\Omega\subset\mathbb{R}^2$ of length $L$ and width $H$, and the coordinate system shown in Figure~\ref{fig:domain}(b).  A pre-crack of length $\xi_0\ll L$ is present on the left as shown.  The boundary $\partial\Omega$ of this domain is subjected to a time-dependent displacement boundary condition 
\begin{equation}
\label{eqn:surfing}
\underline{u}_0(\underline{z},t)=\underline{U}_0(\underline{z}-Vt \,\underline{e}_\text{x})
\end{equation}
where $\underline{U}_0$ is some displacement field that opens the crack.  Specifically, $\underline{U}_0$ is taken to the asymptotic  displacement associated with a mode-I crack~\citep{Zehnder-2012},
\begin{equation}
\label{eqn:U0}
\underline{U}_0= \psi \frac{K_{\text{Ic}}(1+\nu)}{E}
\left(\kappa-\cos\phi \right)
\sqrt{\frac{r}{2\pi}}\left(\cos\frac{\phi}{2}\underline{e}_\text{x}+
\sin\frac{\phi}{2}\underline{e}_\text{y}\right)\,,\quad 
\text{with}
\quad
\kappa=
\left\{
\begin{aligned}
&\frac{3-\nu}{1+\nu}  \hspace{1em}  \text{if plane stress}\\
&3-4\nu \hspace{0.8em} \text{if plane strain} 
\end{aligned}
\right.
\end{equation}
where $K_{\text{Ic}}=\sqrt{E' G_\text{c}}$ denotes the critical value of the stress-intensity factor in plane stress or plane strain conditions, and $\psi$ is an arbitrary non-dimensional scaling parameter.   Note that the boundary condition  (\ref{eqn:surfing}) seeks to drive the crack at a steady velocity $V$ at the macroscopic scale.

The actual propagation of the crack as well as the evolution of the elastic and plastic fields in the domain is computed using both the numerical method described earlier.  Unless otherwise stated, the Young modulus $E=1$, Poisson ratio $\nu=0.2$, toughness $G_\text{c}=1$, regularization length  $\ell=0.25$, mesh size $\delta=0.4\ell$, the von Mises strength 
$\sigma_0 =0.5$ (corresponding to a ductility ratio $r_y = 2.5$), and the scaling factor $\psi=1$.  Both minimization procedures \texttt{AUP} and \texttt{UPA} yield similar results.

%\begin{figure}
%\centering
%   {\includegraphics[width=1\textwidth]{Figure_energies_angles_10_20.jpg}}\\
%   \hspace{0.5em}
%   {\includegraphics[width=1\textwidth]{Figure_energies_angles_30_40.jpg}}\\
%   \hspace{0.5em}
%   {\includegraphics[width=1\textwidth]{Figure_energies_angle_80.jpg}}
%\caption{
%{{\color{red}Crack nucleation at a V-notch in plane strain and stress conditions. Elastic-plastic material $\mathcal{M}_\text{ep}$ for different values of the notch angle $\omb $. Energies as a function of time $t$, for $r_\text{y}=2.0$.}}}
%\label{fig:PACMAN_energies_angles}
%\end{figure}

%%%%%%%%%%%%%%%%%%%%%%%%%%%%%%%%%%%%%%%%%%%%%%%%%%%%%
%%%%%%%%%%%%%%%%%%%%%%%%%%%%%%%%%%%%%%%%%%%%%%%%%%%%%
\subsection{Crack propagation}

\begin{figure}
\centering
\includegraphics[width=1\textwidth]{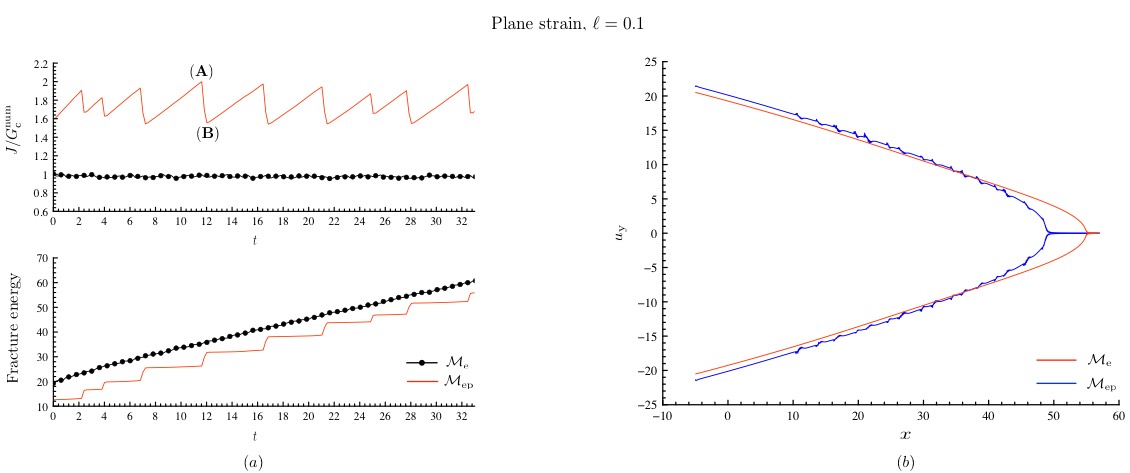} \\ \vspace{0.1in}
\includegraphics[width=1\textwidth]{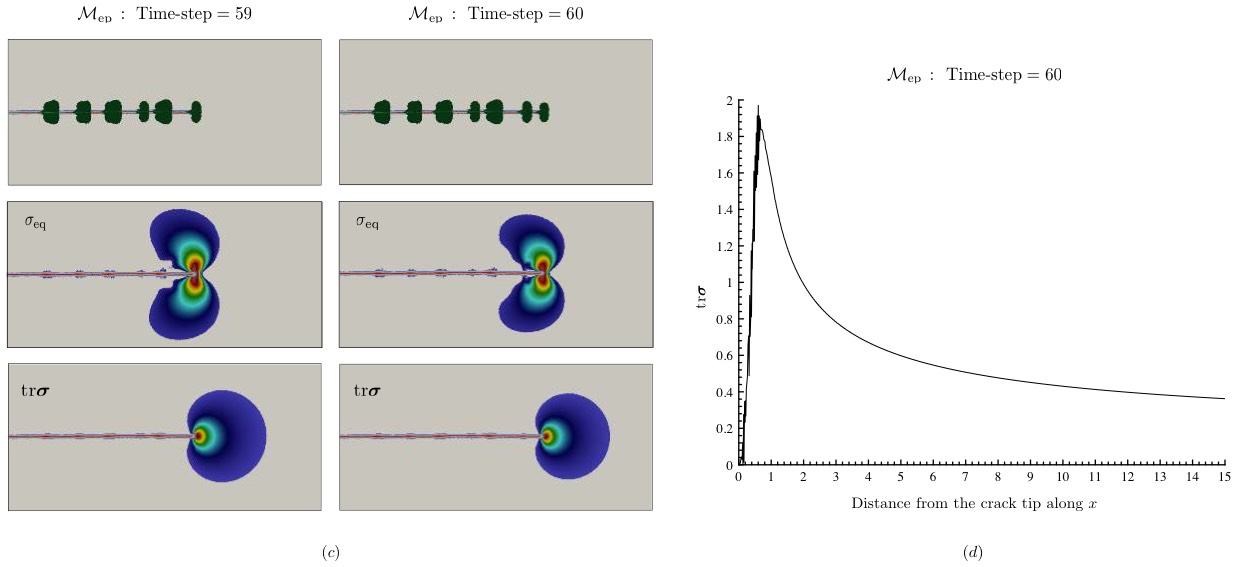} \\
\caption{
{Crack propagation in plane strain conditions. Comparison between an elastic $\mathcal{M}_\text{e}$ and elastic-plastic $\mathcal{M}_\text{ep}$ material, for $\ell=0.1$. (a) $J$-integral and fracture energy as a function of the time parameter $t$. (b) Fracture surface computed through the opening displacement $u_\text{y}$ for a time-step equal to $190$. (c) Plastic strain and stress fields distributions. Thresholds on the hydrostatic and equivalent stress measures are respectively set equal to $0.5$ and $0.15$. (d) Hydrostatic stress as a function of the distance from the crack tip along the $x$-axis.}}
\label{fig:HOM_J_time}
\end{figure}

\paragraph{Plane Strain.}
Figure~\ref{fig:HOM_J_time} shows the typical result for a ductile material in {\it plane strain} conditions and contrasts it with that of an elastic material.  Initially, the energy release rate of the applied boundary condition is too small to drive the crack and it remains stationary.  As time passes and the center of the applied macroscopic crack tip $Vt$ moves to the right, the energy release rate increases with no crack growth.  A plastic zone consisting of two lobes appear at the crack-tip.  At some point, the energy release rate reaches a critical value and the crack jumps in the elastic-plastic material, and this is accompanied by a drop in the energy release rate.  The crack is arrested and the plastic zone grows and energy release rate increases with continued loading till it reaches a critical value.  The crack then jumps and the cycle repeats.  In other words, crack propagation in the elastic-plastic material follows a jerky or intermittent pattern of alternating arrest and jump despite the fact that the macroscopic driving is steady.  A consequence of the jerky motion is that the plastic zone is non-uniform and provides a record of the locations where the crack is arrested.  This results in a crack surface that is rough as is clear from the crack opening profile shown in Figure~\ref{fig:HOM_J_time}(b).  It has been verified (see Supplementary material) that these results do not depend on the length of the pre-crack ($\xi_0$) and the scaling factor $\psi$ of the applied boundary condition.  

This jerky evolution of the ductile material is different from the relatively steady evolution that is observed in the purely elastic (brittle) case also shown in Figure~\ref{fig:HOM_J_time}.  Further, the effective toughness (i.e., the peak applied energy release rate) of the elastic-plastic (ductile) material is significantly higher than that of the elastic (brittle) material.

It has long been recognized that the ductile material would be tougher than the brittle material due to the extra dissipation of plasticity (see~\citep{h_book_79} and references there).  However, the observation of jerky motion appears new, and a departure from the arguments of~\citep{Rice-1968,rs_jmps_78, h_book_79}.  They compare a pinned crack (in an infinite medium loaded at infinity) and one that is steadily propagating, and show that the former has a smaller incremental energy release rate than that of the latter.  They argue if the pinned crack were to start propagating continuously with time, the energy release rate would increase monotonically thereby ensuring the incremental stability of this solution.  The results here show that there is another solution to the problem -- one where the crack jumps.  The setting of rate independent, perfect plasticity means that one can have multiple solutions.  Further incremental stability of a solution involving crack propagating continuously with time does not imply stability against a solution involving crack jumps.

The source of the intermittent crack growth is evident by examining the hydrostatic stress, also shown in Figure~\ref{fig:HOM_J_time} (c,d).  This is positive and reaches its peak ahead of the crack tip.  This would suggest that a daughter crack would nucleate ahead of the crack tip and then propagate backwards to join the main crack.  This happens instantaneously in the current rate-independent, perfectly plastic model and manifests itself as a crack jump.   This understanding is consistent with the recent work cited in~\citep{bl_aam_10}  (also~\citep{betal_am_04,petal_am_16}) who argue that voids would nucleate ahead of the crack tip and eventually coalesce.  Further, the experimental observations of~\citep{kg_jmps_89} clearly shows the appearance of a daughter crack ahead of the main crack followed by coalescence albeit in a dynamical setting.  Finally, the jerky motion results in a rough or dimpled crack surface and this is well established in fractography~\citep{h_book_99}.

\begin{figure}
\centering
   {\includegraphics[width=1\textwidth]{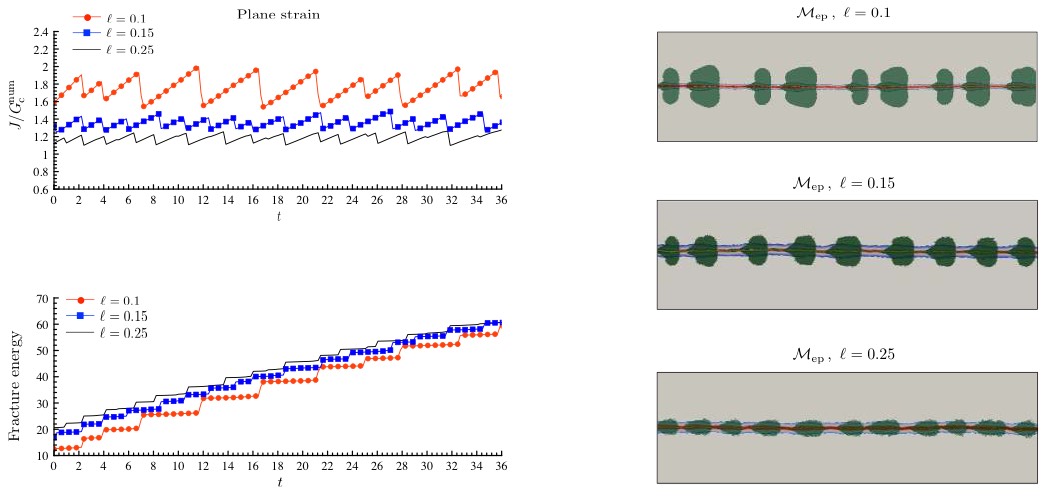}}\\
   {\includegraphics[scale=0.5]{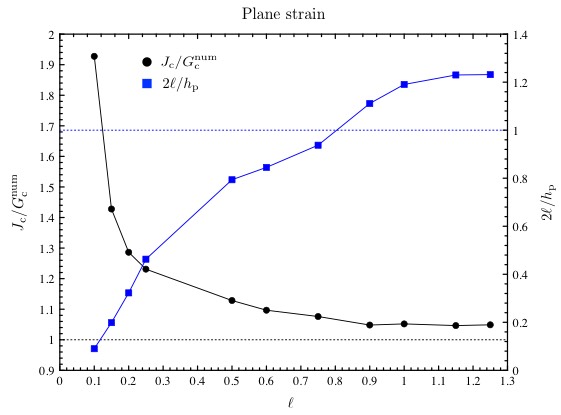}}
\caption{
{Crack propagation in plane strain conditions.  Influence of the regularization parameter $\ell$.   The mesh size $\delta$ is correspondingly varied to keep a constant ratio $\delta/\ell=0.4$; so the the numerical toughness $G_\text{c}^\text{num}$ in elastic materials is constant.}}
\label{fig:HOM_ell}
\end{figure}

\begin{figure}
\centering
   {\includegraphics[width=1\textwidth]{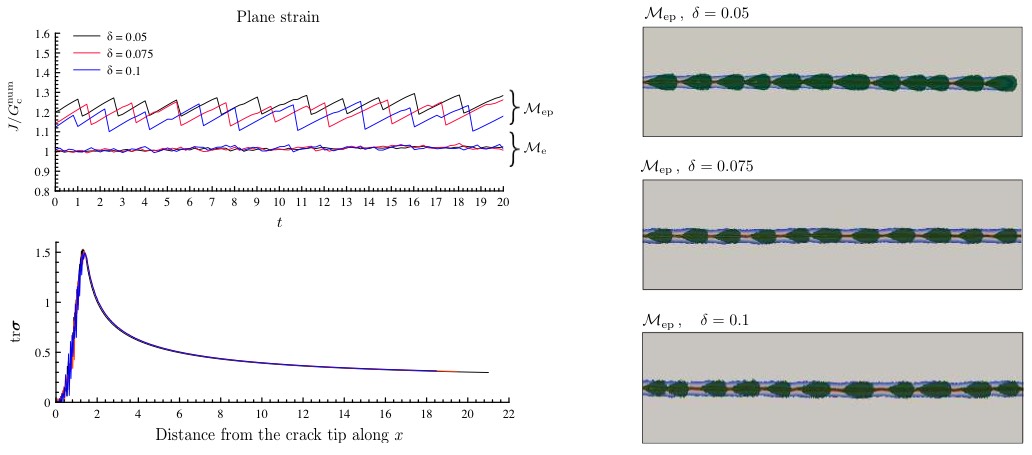}}
\caption{
{Crack propagation in plane strain conditions. Influence of the mesh size $\delta$ for a constant value of the regularization parameter $\ell=0.25$. }}
\label{fig:HOM_mesh}
\end{figure}

The role of regularization parameter $\ell$ is studied in Figure~\ref{fig:HOM_ell} holding the ratio $\delta/\ell$ and the yield strength $\sigma_0$ constant. This keeps the numerical toughness $G_\text{c}^\text{num}$ in elastic materials the same as $\ell$ is varied. However, notice that increasing $\ell$ decreases the nucleation stress  $\sigma_\text{c}$ (see Eq.\,\eqref{eqn:nucleation}) which in turn makes the material less ductile by decreasing $r_y$  (see Eq.\,\eqref{eq:ry}).  Thus, the effective toughness of elastic-plastic materials $J_c$ (peak value of $J$) and the thickness $h_\text{p}$ of the plastic process zone decrease as shown in the figure.

Finally, the mesh size  $\delta$ influences the crack propagation in two ways as shown in Figure~\ref{fig:HOM_mesh}.  Recall from (\ref{eq:gcnum}) that the numerical toughness in the purely brittle case $G_c^\text{num}$ depends on $\delta$.  This is normalized in the figure as is clear from the purely elastic case.  Still the effective toughness given by the peak energy release rate is affected by $\delta$.  This suggests that the discretization affects the effective toughness differently in the ductile and brittle cases.  However, the stress and distribution as well as the jerky propagation of the cracks is not affected by the discretization.  

\begin{figure}
\centering
   {\includegraphics[scale=0.5]{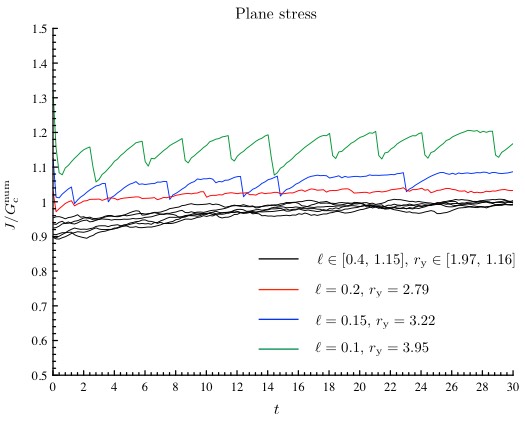}}
\caption{
{Crack propagation in plane stress.  The macroscopic $J$-integral as a function of time for various values of regularization length $\ell$. The mesh size $\delta$ is correspondingly varied such that the numerical toughness $G_\text{c}^\text{num}$ in elastic materials is constant. }}
\label{fig:HOM_plane_stress}
\end{figure}

\begin{figure}
\centering
   {\includegraphics[width=1\textwidth]{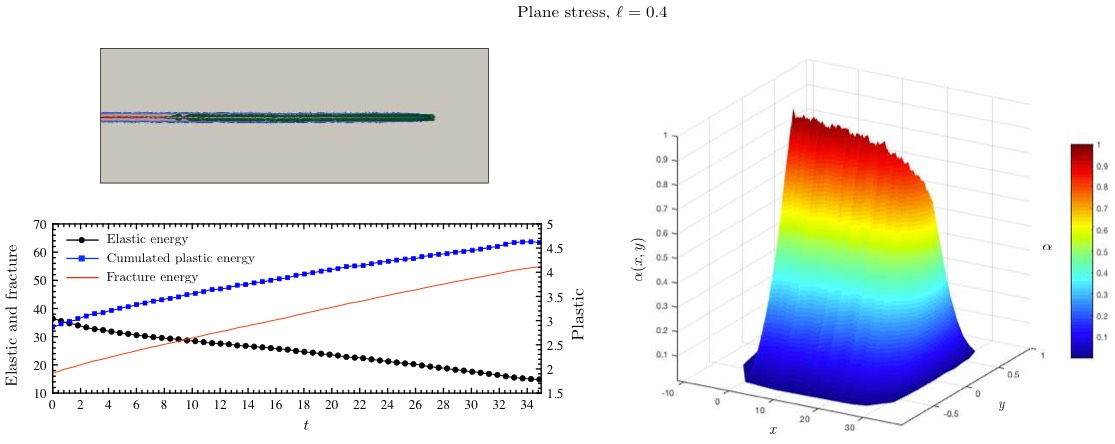}}
\caption{
{Crack propagation in plane stress for a material with small ductility ratio ($r_y = 1.97$ corresponding to $\ell=0.4$).}}
\label{fig:HOM_plane_stress_ell_04}
\end{figure}

\begin{figure}
\centering
   {\includegraphics[width=1\textwidth]{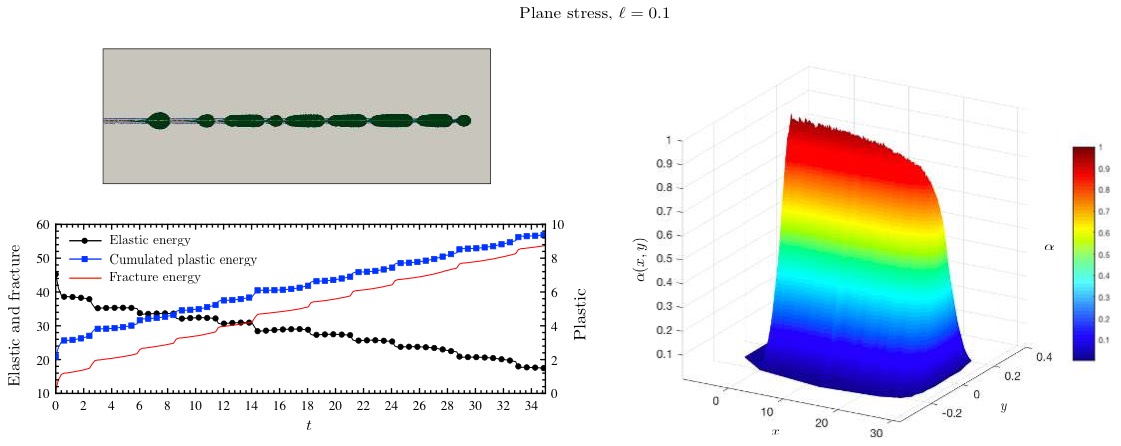}}
\caption{
{Crack propagation in plane stress for a material with large ductility ratio ($r_y = 3.95$ corresponding to $\ell=0.1$).}}
\label{fig:HOM_plane_stress_ell_01}
\end{figure}

\paragraph{Plane stress.} Figure~\ref{fig:HOM_plane_stress} collects the results for a variety of elastic-plastic materials with varying regularization length $\ell$ but constant von Mises strength $\sigma_0$ (so the ductility ratio $r_y$ changes) and numerical toughness $G_\text{c}^\text{num}$ (i.e., $\delta$ is changed to keep the ratio $\delta$/$\ell$ constant).  Figure~\ref{fig:HOM_plane_stress_ell_04} shows more details for a representative material with small ductility ratio $r_y$ and Figure~\ref{fig:HOM_plane_stress_ell_01} for a representative material with large ductility ratio.

For small values of $r_\text{y}$, a region of intense plastic deformation and fracture process develops at the pre-crack tip and begins to grow steadily when the energy-release rate reaches a critical value, just below the toughness $G_c^\text{num}$.  Apart from numerical oscillations, the measured $J$-integral remains essentially constant as this zone grows.  The fracture field in this region is close to but not exactly equal to the one; so the fracture process is not complete.  
A different scenario arises when $r_\text{y}$ increases: the plastic process zone becomes larger and starts interacting with fracture in a more complicated and jerky manner resembling crack growth in the plane strain conditions.  However, again, the fracture process is not complete.

\paragraph{Summary.}

Crack propagation depends strongly on the ratio $r_\text{y}$ between the nucleation stress and the yield strength.  In the quasi-brittle situation where this ratio is small ($r_y <1$), the crack propagation is similar to that in purely elastic materials -- it is steady and the toughness is close to the prescribed value $G_c^\text{num}$.  The ductile situation $r_y  > 1$ is different.  In plane strain, crack propagation proceeds in a jerky fashion with a plastic zone pinning the crack till the imposed driving force reaches a high value when a crack advances with a finite jump, gets re-pinned and the cycle repeats.  This mechanism is a result of the fact that the stress triaxiality is highest ahead of a pinned crack.  The effective toughness is higher than the prescribed value $G_c^\text{num}$, and increases with ductility ratio $r_y$.  The plastic deformation is not uniform and the fracture surface is rough.   In plane stress, failure is accompanied by intense plastic deformation ahead of the crack tip and the fracture process is not complete.

%\begin{figure}
%\centering
% \subfloat[]
%   {\includegraphics[width=1\textwidth]{Figure18a.jpg}}\\
%   \subfloat[]
%   {\includegraphics[width=1\textwidth]{Figure18b.jpg}}
%\caption{
%{Crack propagation: energies and $J$-integral as a function of the time variable for an elastic-plastic material $\mathcal{M}_\text{ep}$ with: (a) $\ell=0.1$ in plane strain (i.e., $r_\text{y}=3.87$) and plane stress (i.e., $r_\text{y}=3.95$) conditions; (b) $\ell=0.05$ in plane stress conditions (i.e., $r_\text{y}=5.59$).  {\color{blue}  Separate this into two figures.   First, with (a) right and (b) focussing on plane stress.  In fact, combine this with the previous figure.  Second, with (a) left contrasting plane stress and strain. } }}
%\label{fig:HOM_confronto_en}
%\end{figure}

\section{Conclusion}\label{sec:concl}

In this paper, crack nucleation and propagation in elastic - perfectly plastic materials has been investigated in a phase-field or variational fracture field setting.  The model proposed by~\citep{Alessi-2014,Alessi-2015} has been implemented numerically and used to study crack nucleation at a sharp notch, crack initiation in a notched specimen and crack propagation under a steady driving.  In a quasi-brittle setting (characterized by the ratio of the critical stress for crack nucleation to yield stress being smaller than unity), plastic deformation is limited to a small region near the crack tip and the nucleation, propagation and toughness is similar to that of brittle (purely elastic) materials.  This is consistent with the notion of small scale yielding~\citep{h_book_79}.

The behavior is different in the ductile setting when the critical stress for nucleation exceeds the yield strength.  In plane strain, there are two symmetric lobes of plastic deformation at the crack (notch) top.  This pins the crack and one needs a very large driving force for the crack to un-pin (nucleate) the crack.  When it does, the crack jumps ahead by a finite length.  There is plastic deformation at the new crack tip and the cycle repeats.  Consequently, the effective toughness of the material is higher, the crack propagation is intermittent or jerky and the fracture surface is rough.  The source of the crack jump appears to be the fact the hydrostatic tension is high ahead of the crack tip.  So, one expects voids to nucleate ahead of the pinned crack tip and then coalesce with the main crack.  This appears instantaneous in our rate-independent, perfectly plastic setting.  In plane stress, failure proceeds by intense plastic deformation ahead of the crack.  The fracture process is not complete.

\section*{Acknowledgments}
We have greatly benefited from many discussions with G. ``Ravi'' Ravichandran, Katherine Faber, Amine Benzerga and Alan Needleman.  SB, BB, and KB acknowledge the financial support of the U.S. National Science Foundation (Grant No. DMS-1535083 and 1535076) under the Designing Materials to Revolutionize and Engineer our Future (DMREF) Program. Some numerical experiments were performed using resources of the Extreme Science and Engineering Discovery Environment (XSEDE), which is supported by National Science Foundation grant number OCI-1053575 under the Resource Allocation TG-DMS060014 and others at the Caltech high performance cluster supported in part by the Moore Foundation.

\bibliography{BTBB_2019}

\newpage

\section*{Supplementary material}
\renewcommand{\thefigure}{S\arabic{figure}}
\setcounter{figure}{0}

\begin{figure}[h!]
\centering
   {\includegraphics[width=0.6\textwidth]{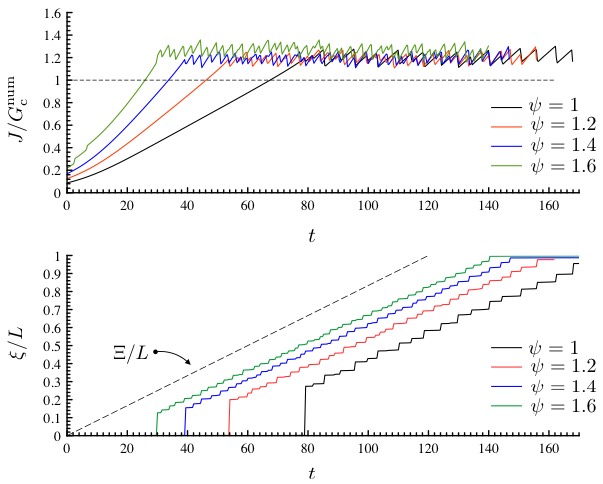}}
\caption{
{Crack propagation in plane strain conditions. $J$-integral and crack length $\xi$ as a function of the time parameter $t$.  Influence of the magnitude $\psi$ of the applied opening displacement in Eq.\,\eqref{eqn:surfing}.  }}
\label{fig:HOM_J_surfing}
\end{figure}

\begin{figure}[btp]
\centering
 \subfloat[]
   {\includegraphics[width=0.9\textwidth]{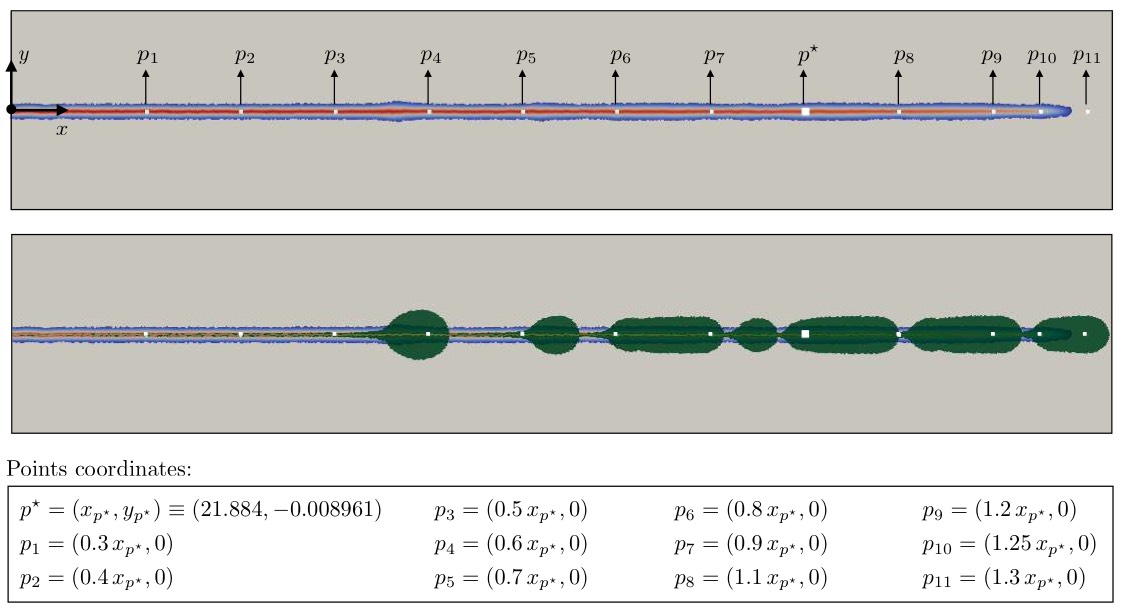}}\\
   \subfloat[]
   {\includegraphics[width=0.9\textwidth]{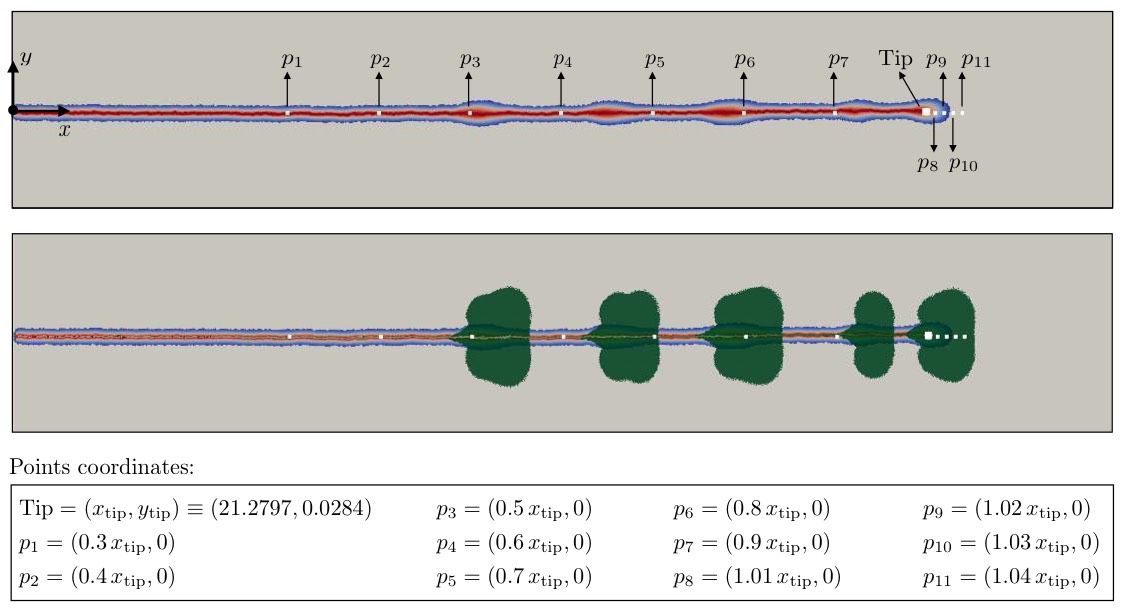}}
\caption{
{Crack propagation for $\ell=0.1$ in (a) plane stress (i.e., $r_\text{y}=3.95$) and (b) plane strain (i.e., $r_\text{y}=3.87$) conditions.  Damage field and equivalent plastic strain. In both cases the time-step is set equal to $50$. Material points $(x,y)$ such that $x>x_{p^\star}$ exhibit values of damage lower than $0.9$.}}
\label{fig:HOM_cracks}
\end{figure}

\begin{figure}[btp]
\centering
 \subfloat[]
   {\includegraphics[width=1\textwidth]{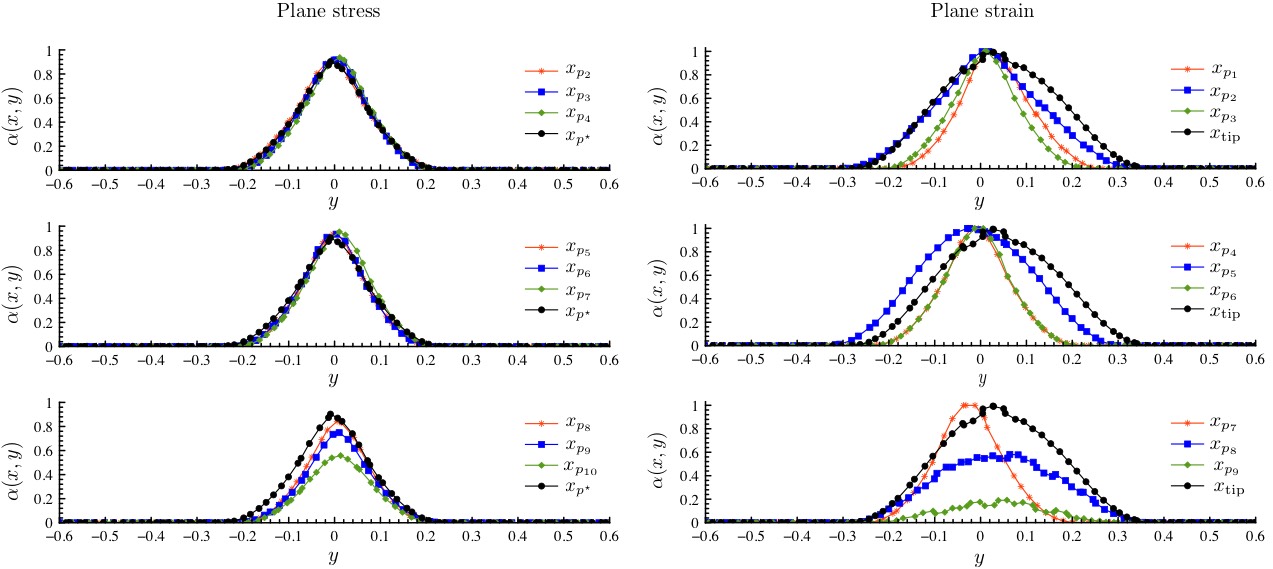}}\\
   \subfloat[]
   {\includegraphics[width=1\textwidth]{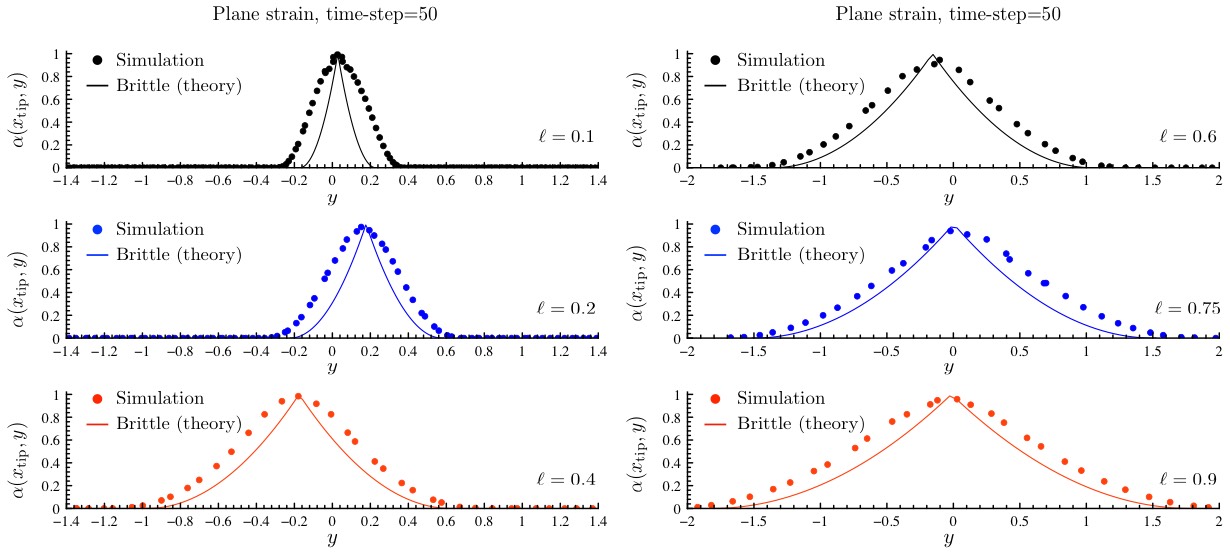}}
\caption{
{Fracture field (see Figs.\,\ref{fig:HOM_cracks} for notation): (a) profiles computed at different coordinates along the $x$-axis, for a time-step equal to $50$.; (b) comparison between theoretical~\citep{Ambrosio-1990} and numerical damage profiles, for different values of the regularization length $\ell$.}}
\label{fig:HOM_num_profiles}
\end{figure}

\end{document}